\documentclass[12pt, preprint,numberedappendix]{emulateapj}

\newcommand\bibinc{n}		

\usepackage{subeqnarray}
\usepackage{amsmath}
\usepackage{hyperref}
\usepackage[normalem]{ulem}

\DeclareMathSymbol{\varOmega}{\mathord}{letters}{"0A}
\DeclareMathSymbol{\varSigma}{\mathord}{letters}{"06}
\DeclareMathSymbol{\varPsi}{\mathord}{letters}{"09}

\newcommand{\Eq}[1]{Equation\,(\ref{#1})}

\newcommand{\App}[1]{Appendix~\ref{#1}}
\newcommand{\Sec}[1]{Section~\ref{#1}}

\newcommand{\Fig}[1]{Figure~\ref{#1}}

\newcommand{\Tab}[1]{Table \ref{#1}}

\newcommand{\twid}{\tilde}

\begin{document}

\slugcomment{Accepted at ApJ}

\shorttitle{Water partitioning on terrestrial exoplanets}
\shortauthors{Komacek \& Abbot}

\title{Effect of surface-mantle water exchange parameterizations on exoplanet ocean depths}
\author{Thaddeus D. Komacek$^1$ \& Dorian S. Abbot$^2$}
\affil{$^1$Lunar and Planetary Laboratory, Department of Planetary Sciences, University of Arizona \\
$^2$Department of the Geophysical Sciences, University of Chicago}
\begin{abstract}
Terrestrial exoplanets in the canonical habitable zone may have a variety of initial water fractions due to random volatile delivery by planetesimals. If the total planetary water complement is high, the entire surface may be covered in water, forming a ``waterworld.'' 
On a planet with active tectonics, competing mechanisms act to regulate the abundance of water on the surface by determining the partitioning of water between interior and surface. Here we explore how the incorporation of different mechanisms for the degassing and regassing of water changes the volatile evolution of a planet. 
For all of the models considered, volatile cycling reaches an approximate steady-state after $\sim 2 \ \mathrm{Gyr}$. Using these steady-states, we find that if volatile cycling is either solely dependent on temperature or seafloor pressure, exoplanets require a high abundance ($\gtrsim 0.3\%$ of total mass) of water to have fully inundated surfaces. 
However, if degassing is more dependent on seafloor pressure and regassing mainly dependent on mantle temperature, the degassing rate is relatively large at late times and a steady-state between degassing and regassing is reached with a substantial surface water fraction. If this hybrid model is physical, super-Earths with a total water fraction similar to that of the Earth can become waterworlds. 
As a result, further understanding of the processes that drive volatile cycling on terrestrial planets is needed to determine the water fraction at which they are likely to become waterworlds.
\end{abstract}
\section{Introduction}
\subsection{Surface water abundance and habitability}
\indent To date, the suite of observed exoplanets from \textit{Kepler} has proven that Earth-sized planets are common in the universe ($\approx 0.16$ per star, \citealp{Fressin2013,Morton2014}). Though we do not yet have a detailed understanding of the atmospheric composition of an extrasolar terrestrial planet, spectra of many extrasolar gas giants \citep{Kreidberg2015a,Sing2015} and a smaller Neptune-sized planet \citep{Fraine2014} have shown that water is likely abundant in other Solar Systems. Calculations of volatile delivery rates to terrestrial planets via planetesimals (e.g. \citealp{Raymond:2004,Ciesla2015}) have shown that planets can have a wide range of initial water fractions, with some planets being $1\%$ water by mass or more. Both observations and simulations hence point towards the likelihood that terrestrial planets are also born with abundant water. However, the intertwined effects of climate \citep{Kasting1993} and mantle-surface volatile interchange \citep{Hirschmann2006,Cowan2015} determine whether there is abundant liquid water on the present-day surfaces of terrestrial exoplanets. Additionally, atmospheric escape (especially early in the atmospheric evolution) can cause loss of copious amounts of water \citep{Ramirez2014,Luger2015,Tian2015,Schaefer2016}, with $\gtrsim 10$ Earth oceans possibly lost from planets in the habitable zone of M-dwarfs.  \\
\indent The extent of the traditional habitable zone is determined by the continental silicate weathering thermostat \citep{Kasting1988}, in which silicate minerals react with $\mathrm{C}\mathrm{O}_2$ and rainwater to produce carbonates \citep{Walker1981}. Silicate weathering is extremely efficient at stabilizing the climate because the process runs faster with increasing temperature. This is due to faster reaction rates and increased rain in warmer climates. However, the silicate weathering thermostat itself depends on the surface water abundance. \\
\indent If there is no surface water, the silicate weathering thermostat cannot operate due to the lack of reactants, and if the planet surface is completely water-covered the negative feedback does not operate unless seafloor weathering is also temperature dependent \citep{Abbot2012}. Note that even if seafloor weathering is temperature-dependent, it might be insufficient to stabilize the climate \citep{Foley2015}. A waterworld state is likely stable \citep{Wordsworth2013}, as water loss rates would be low because the atmosphere would be $\mathrm{C}\mathrm{O}_2$-rich due to the lack of a silicate-weathering feedback. However, if water loss rates remain high due to a large incident stellar flux, it is possible that brief exposures of land can allow for a ``waterworld self-arrest'' process in which the planet adjusts out of the moist greenhouse state \citep{Abbot2012}. This can occur if the timescale for $\mathrm{C}\mathrm{O}_2$ drawdown by the silicate-weathering feedback is shorter than the timescale for water loss to space, which is probable for Earth parameters. \\
\indent From the above discussion, we conclude that although waterworlds are by definition in the habitable zone (having liquid water on the surface), they may not actually be temperate and conducive to life. It is instead likely that waterworlds are less habitable than worlds with continents, and so determining whether or not waterworlds are common is important. To determine whether or not waterworlds should be common, we must look to the deep-water cycle, that is, the mantle-surface interchange of water over geologic time. 
\subsection{Earth's deep-water cycle}
\label{sec:deepintro}
\indent To understand the deep-water cycle on exoplanets, we look to Earth as an analogue, as it is the only planet known with continuous (not episodic) mantle-surface water interchange due to plate tectonics. On present-day Earth, water is largely expelled from the mantle to the surface (degassed) through volcanism at mid-ocean ridges and volcanic arcs \citep{Hirschmann2006}. Water is lost from the surface to the mantle (regassed) through subduction of hydrated basalt. The relative strength of regassing and degassing determines whether the surface water abundance increases or decreases with time.  \\
\indent It has long been suggested that Earth's surface water fraction is in effective steady-state \citep{McGovern1989,Kasting1992}, due to the constancy of continental freeboard since the Archean ($\sim 2.5$ Gya). However, this may simply be due to isostasy, that is, the adjustment of the continental freeboard under varying surface loads \citep{Rowley2013,Cowan2014a}. A more convincing argument is that the degassing and regassing rates on Earth are high enough that if they did not nearly balance each other the surface would have long ago become either completely dry or water-covered \citep{Cowan2014a}. However, some studies of volatile cycling on Earth that utilized parameterized convection to determine the upper mantle temperature and hence the degassing and regassing rates have not found such a steady state \citep{McGovern1989,Crowley2011,Sandu2011}. If the Earth is indeed near steady-state, this mismatch could be because there are many secondary processes, e.g. loss of water into the transition zone \citep{Pearson2014} and early mantle degassing \citep{Elkins-Tanton2011}, that are difficult to incorporate into a simplified volatile cycling model. 
Also, it is possible that our understanding of what processes control the release of water from the mantle and return of water to it via subduction is incomplete.  \\
\indent Using the maximum allowed fraction of water in mantle minerals \citep{Hauri2006,Inoue2010}, \cite{Cowan2014a} estimate that Earth's mantle water capacity is $\approx 12$ times the current surface water mass. However, measurements of the electrical conductivity of Earth's mantle \citep{Dai2009} have found only $\sim 1-2$ ocean masses of water in the mantle, which is much less than the maximally allowed value. This measurement may vary spatially \citep{Huang2005} and by method \citep{Khan2012}, but it is likely constrained to within a factor of a few. This implies that dynamic effects lead to a first-order balance between degassing and regassing on Earth, rather than the surface water complement being in steady-state simply because the mantle is saturated.
\subsection{Previous work: the deep-water cycle on super-Earths}
\indent Using a steady-state model wherein the degassing and regassing of water is regulated by seafloor pressure, \cite{Cowan2014a} applied our knowledge of Earth's deep-water cycle to terrestrial exoplanets. They showed that terrestrial exoplanets require large amounts ($\sim 1\%$ by mass) of delivered water to become waterworlds. Applying a time-dependent model and including the effects of mantle convection, \cite{Schaefer:2015} found that the amount of surface water is strongly dependent on the details of the convection parameterization. These works rely on other planets being in a plate-tectonic regime similar to Earth. However, it is important to note that there is debate about whether or not plate tectonics is a typical outcome of planetary evolution (e.g. \citealp{ONeill2007,Valencia2007,Valencia2009,Korenaga2010}), potentially because plate tectonics is a history-dependent phenomenon \citep{Lenardic2012}. In this work, we also assume plate tectonics. We do so because our understanding of habitability is most informed by Earth and it enables us to examine how processes that are known to occur on Earth affect water cycling on exoplanets. As a result, we assume that continents are present, and that isostasy determines the depths of ocean basins. In the future, exploring other tectonic regimes (e.g. stagnant lid) may be of interest to exoplanet studies and potential investigations of Earth's future evolution \citep{Sleep2015}. \\
\indent The studies of volatile cycling on super-Earths discussed above used drastically different approaches, with \cite{Cowan2014a} applying a two-box steady-state model of volatile cycling, and \cite{Schaefer:2015} extending the time-dependent coupled volatile cycling-mantle convection model of \cite{Sandu2011} to exoplanets. As a result, these works made different assumptions about which processes control water partitioning between ocean and mantle. The degassing parameterization of \cite{Cowan2014a}, based on the model of \cite{Kite2009a}, utilized the negative feedback between surface water inventory and volatile degassing rate that results from pressure reducing degassing. Their regassing rate was also related to the surface water inventory, using the prediction of \cite{Kasting1992} that the hydration depth increases with increasing surface water abundance up to the limit where the hydration depth is equal to the crustal thickness. Meanwhile, the degassing and regassing parameterizations of \cite{Schaefer:2015} were both related directly to the mantle temperature, with the degassing rate determined by the abundance of water in melt and the regassing rate set by the depth of the hydrated basalt (serpentinized) layer, which is determined by the depth at which the temperature reaches the serpentinization temperature.  \\
\indent In this work, we seek to identify how different assumptions about regassing and degassing determine the surface water mass fraction. To do so, we utilize simplified models of convection and volatile cycling that separately incorporate the key features of both the \cite{Cowan2014a} and \cite{Schaefer:2015} volatile cycling parameterizations. The latter model builds upon the analytic work of \cite{Crowley2011}, who developed an analytic model that captures the key processes in the numerical models of \cite{Sandu2011} and \cite{Schaefer:2015}. However, here we further simplify and also non-dimensionalize the \cite{Crowley2011} model, enabling us to elucidate the dependencies of water abundance on mantle temperature and planetary parameters. We then combine the models of \cite{Cowan2014a} and \cite{Schaefer:2015}, utilizing surface water budget-dependent degassing and temperature-dependent regassing. We do so because it is likely the most physically relevant choice, as temperature affects serpentinization depths (and resulting regassing rates) more directly than seafloor pressure. Additionally, temperature-dependent degassing would become small at late times while seafloor pressure-dependent degassing would not, and it has been shown by \cite{Kite2009a} that degassing should be pressure-dependent. This is more in line with the approximate steady-state water cycling on Earth is currently in, as if both regassing and degassing are temperature-dependent regassing will dominate at late times. We find that the choice of volatile cycling parameterization greatly impacts the end-state surface water mass reservoir. We also find that, regardless of volatile cycling parameterization, the water partitioning reaches a steady-state after a few billion years of evolution due to the cooling of the mantle to below the melting temperature, which causes the effective end of temperature-dependent degassing and regassing. \\
\indent This paper is organized as follows. In \Sec{sec:theory}, we describe our parameterized convection model and the various volatile cycling parameterizations we explore, along with the consequences these have for the temporal evolution of mantle temperature and water mass fraction. Detailed derivations of the volatile cycling models can be found in \App{app:deriv}. In \Sec{sec:waterworld} we explore where in water mass fraction-planet mass parameter space each volatile cycling model predicts the waterworld boundary to lie. We discuss our results in \Sec{sec:discussion}, performing a sensitivity analysis of the waterworld boundary on key controlling parameters, comparing this work to previous works, and discussing our limitations and potential avenues for future work. Importantly, we also show how our model with pressure-dependent degassing and temperature-dependent regassing could in principle be observationally distinguished from the models of \cite{Cowan2014a} and \cite{Schaefer:2015}. Lastly, we express conclusions in \Sec{sec:conc}.
\section{Coupling mantle convection and volatile cycling}
\label{sec:theory}
\subsection{Parameterized convection}
\label{sec:convbackground}
\begin{figure}
	\centering
	\includegraphics[width=.5\textwidth]{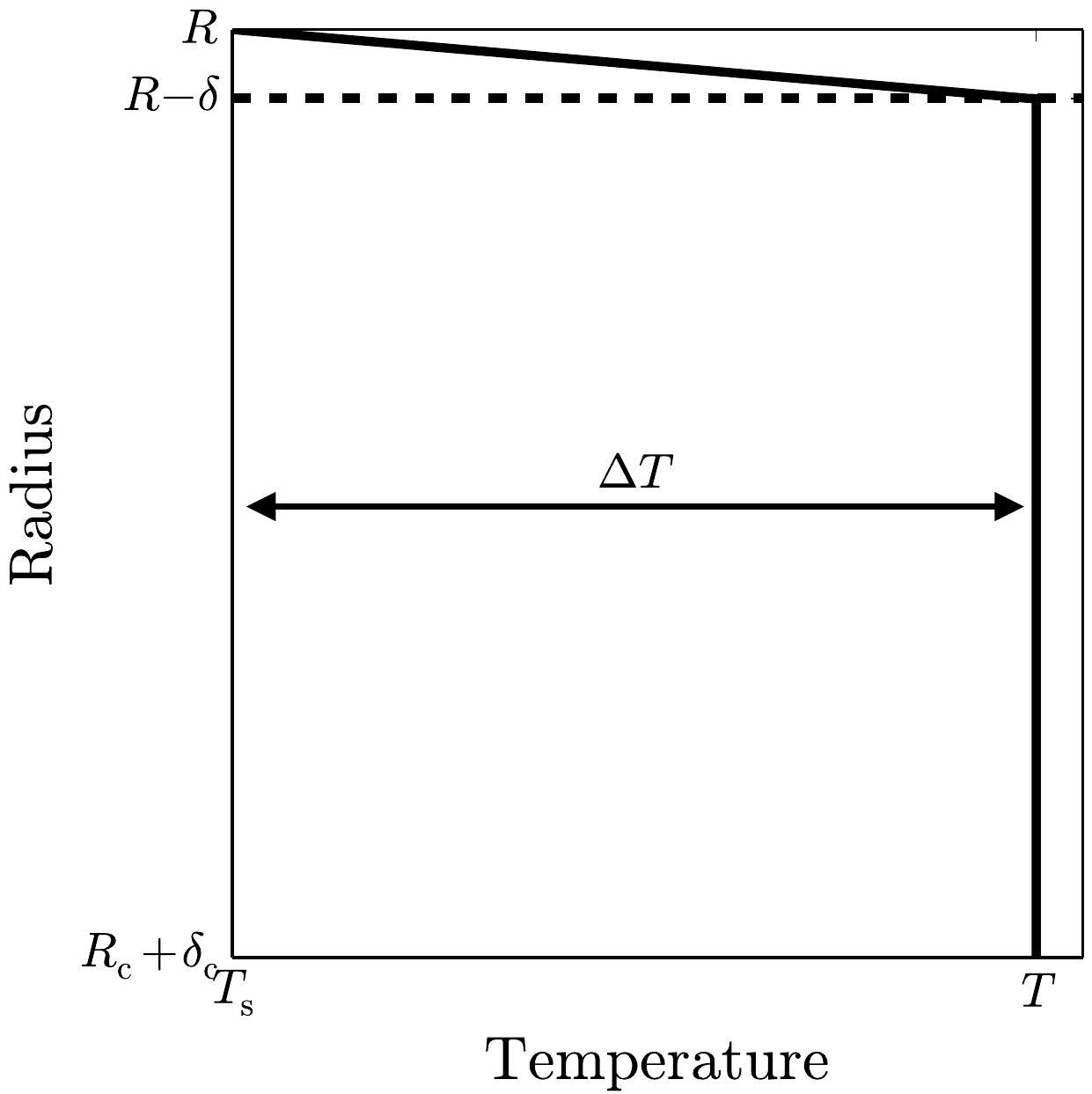}
	\caption{Schematic of the temperature profile utilized for the parameterized convection model. The mantle temperature, $T$ is constant with depth throughout the convecting mantle. A conducting boundary layer forms at the top (above the dashed line) of this convecting interior, of thickness $\delta$. This boundary layer has a temperature contrast $\Delta T$ across it. Here $R$ is planet radius, $T_\mathrm{s}$ is surface temperature, and $R_\mathrm{c}+\delta_{\mathrm{c}}$ is the radius at the boundary between the mantle and the lower boundary layer that separates core and mantle. We do not consider the heat transfer between the core and mantle in this model due to the high viscosity of the lower boundary layer and therefore low heat flux into the mantle.}
	\label{fig:schem}
\end{figure}
\indent Parameterized convection models utilize scalings from numerical calculations to relate the Nusselt number (the ratio of outgoing heat flux from the mantle to that which would be conducted if the entire mantle were not convecting) to the Rayleigh number of the mantle \citep{Turcotte:2002}. We here consider a standard boundary-layer convection model, as in \cite{McGovern1989} and \cite{Sandu2011}, with a top boundary layer of thickness $\delta$ and one characteristic mantle temperature $T$. \Fig{fig:schem} shows a schematic of the temperature profile relevant for this convection parameterization. We can determine the boundary-layer thickness by the depth at which the boundary layer peels away and convects. To zeroth order, this peel-off occurs where the timescale for advection of the boundary layer is shorter than the timescale for heat to diffuse out of the boundary layer. The time it takes for the boundary-layer to overturn via advection is 
\begin{equation}
\label{eq:tauoverturn}
\tau_{\mathrm{over}} \approx \frac{\eta(T,x)}{\Delta \rho g \delta} \mathrm{,}
\end{equation}
where $\eta(T,x)$ is the temperature and mantle water fraction-dependent viscosity of the boundary-layer (the viscosity parameterization will be discussed further in \Sec{sec:viscosity}), $g = g(M)$ is gravity (see \Sec{sec:mass} for how $g$ and other variables scale with planet mass), and we take the density contrast $\Delta \rho_{\mathrm{m}} \approx \alpha \rho \Delta T$, where $\alpha$ is a characteristic thermal expansivity, $\Delta T$ the temperature contrast across the boundary-layer, and $\rho_{\mathrm{m}}$ is the density of the upper mantle. The heat diffusion timescale is then
\begin{equation}
\label{eq:taudiff}
\tau_{\mathrm{diff}} \approx \frac{\delta^2}{\kappa} \mathrm{,}
\end{equation}
where $\kappa$ is the thermal diffusivity of the boundary-layer (assumed equal to that of the upper mantle). Taking the ratio of Equations (\ref{eq:taudiff}) and (\ref{eq:tauoverturn}) defines the local boundary-layer Rayleigh number
\begin{equation}
\label{eq:rayleighloc}
\mathrm{Ra}_{\mathrm{loc}} = \frac{\alpha \rho_{\mathrm{m}} g \Delta T \delta^3}{\eta \kappa} \mathrm{.}
\end{equation}
Then, boundary-layer peel-away occurs when the local Rayleigh number is greater than the critical Rayleigh number for convection (i.e. $\mathrm{Ra}_{\mathrm{loc}} > \mathrm{Ra}_{\mathrm{crit}}$, where $\mathrm{Ra}_{\mathrm{crit}} \sim 1100$). Setting $\mathrm{Ra}_{\mathrm{loc}} = \mathrm{Ra}_{\mathrm{crit}}$, solving for the boundary-layer thickness, and substituting in the mantle Rayleigh number
\begin{equation}
\label{eq:rayleigh}
\mathrm{Ra} = \frac{\alpha \rho_{\mathrm{m}} g \Delta T h^3}{\eta \kappa} \mathrm{,}
\end{equation}
where $h = h(M)$ is the mantle thickness, we find
\begin{equation}
\label{eq:delta}
\delta \sim h \left(\frac{\mathrm{Ra}_{\mathrm{crit}}}{\mathrm{Ra}}\right)^{1/3} \mathrm{.}
\end{equation}
Note that $h$ itself cancels out in \Eq{eq:delta} when inserting in \Eq{eq:rayleigh}, as we have substituted in the mantle Rayleigh number in order to motivate the scaling relationship derived from numerical simulations (see \Eq{eq:flux} below). \\
\indent Using \Eq{eq:delta}, we can find the conducted flux through the boundary layer
\begin{equation}
\label{eq:fluxtheory}
F = \frac{k \Delta T}{\delta} \sim \frac{k \Delta T}{h}\left(\frac{\mathrm{Ra}}{\mathrm{Ra}_{\mathrm{crit}}}\right)^{1/3} \mathrm{,}
\end{equation}
where $k$ is the thermal conductivity of the boundary-layer. In this work, we take a more general power-law form for our Nusselt number scaling which relates the outgoing flux from the mantle to the Rayleigh number:
\begin{equation}
\label{eq:flux}
\frac{F}{F_{\mathrm{cond}}} \equiv \mathrm{Nu} = a \left(\frac{\mathrm{Ra}}{\mathrm{Ra}_{\mathrm{crit}}}\right)^{\beta} \mathrm{.}
\end{equation}
In \Eq{eq:flux}, $\mathrm{Nu}$ is the Nusselt number, $F$ is the convected flux, and $F_{\mathrm{cond}} = k\Delta T/h$ is the flux that would be conducted if the mantle were not convecting. It is expected from numerical studies of convection \citep{schubert79} that $\beta \approx 0.3$, which we take as our nominal value. Note that if $\beta$ is smaller than the value assumed here, planetary thermal evolution would be slower. As in \cite{Schaefer:2015} we set $a = 1$, as $a$ is an order-unity parameter and using our theoretically derived outgoing flux from \Eq{eq:fluxtheory} we expect that $a$ should be equal to one. Note that this model only requires the characteristic temperature at the interface between the upper boundary layer and mantle. As a result, we do not consider the actual (nearly adiabatic) temperature profile of the mantle. Additionally, the temperature contrast across the boundary layer is much greater than that between the top and bottom of the mantle. Given that the argument for convection driven by boundary-layer peel off requires local quantities (e.g. $\kappa$, $\alpha$ relevant for the boundary-layer itself) rather than global quantities, we consider the upper-mantle viscosity $\eta$ in our model. This results in a pressure-independent viscosity, as will be discussed further in \Sec{sec:viscosity}.  \\
\indent Given the flux conducted out of the mantle from \Eq{eq:flux}, we can write down a thermal evolution equation that allows us to solve for the mantle temperature as a function of time and mantle water mass fraction. This is  
\begin{equation}
\label{eq:heat}
\rho_{\mathrm{m}} c_p \frac{dT}{dt} = Q - \frac{A(M) F(T,x)}{V(M)} \mathrm{,}
\end{equation}
where $c_p$ is the mantle heat capacity, $Q = Q_0 e^{-t/\tau_{\mathrm{decay}}}$ is the heating rate from radionuclides with $\tau_{\mathrm{decay}} = 2$ Gyr, $A(M)$ is the planet surface area, and $V(M)$ is the mantle volume. We do not include the Kelvin-Helmholtz contraction term, which is small at late times. We will non-dimensionalize \Eq{eq:heat} in \Sec{sec:nondimtherm} to elucidate its dependence on temperature and mantle water mass fraction.
\subsubsection{Scaling with planet mass}
\label{sec:mass}
To calculate mass-dependent planetary parameters ($h,A,V,g$) we use the scaling laws of \cite{Valencia2006} that take into account internal compression effects on the radius. These scaling relations utilize a constant core mass fraction to relate the planetary radius $R$ and core radius $R_\mathrm{c}$ to planetary mass
\begin{equation}
\label{eq:radiusmass}
\begin{aligned}
& R = R_{\oplus}\left(\frac{M}{M_{\oplus}}\right)^{p} \mathrm{,} \\
& R_\mathrm{c} = cR_{\oplus}\left(\frac{M}{M_{\oplus}}\right)^{p_\mathrm{c}} \mathrm{,}
\end{aligned}
\end{equation} 
where $p = 0.27$, $c = 0.547$, $p_\mathrm{c} = 0.25$. Using \Eq{eq:radiusmass}, we can then calculate $h = R - R_\mathrm{c}$, $A = 4\pi R^2$, $V = 4\pi/3(R^3 - R^3_\mathrm{c})$, $g = GM/R^2$. 
\subsubsection{Viscosity}
\label{sec:viscosity}
\indent The mantle viscosity depends both on temperature and mantle water fraction. 
We use a similar parameterization as \cite{Sandu2011} and \cite{Schaefer:2015} for the mantle viscosity, however, we choose not to incorporate the pressure-dependence of viscosity. As discussed in \Sec{sec:convbackground}, we do so because we are interested in convection driven by upper boundary peel-off, which occurs in the upper mantle where pressures are relatively small. Additionally, the high-viscosity case of the \cite{Schaefer:2015} water cycling model did not reproduce Earth's near steady-state or present ocean coverage. This is because the evolution timescales are too long to reach steady-state in the high viscosity case. However, when utilizing low viscosities, the system does converge to an approximate steady-state surface water mass fraction in all cases.
We show in \Sec{sec:comparison} that this choice of viscosity approximates Earth's mantle temperature well when we choose Earth-like parameters. \\
\indent Our viscosity is hence parameterized as 
\begin{equation}
\label{eq:viscosity}
\eta \approx \eta_0 f_{\mathrm{w}}^{-r} \mathrm{exp}\left[\frac{E_\mathrm{a}}{R_{\mathrm{gas}}}\left(\frac{1}{T}-\frac{1}{T_{\mathrm{ref}}}\right)\right] \mathrm{,}
\end{equation}
where $\eta_0$ gives the viscosity scale, $E_\mathrm{a}$ is activation energy, $R_{\mathrm{gas}}$ is the universal gas constant, $T_{\mathrm{ref}}$ is the reference mantle temperature, and $f_\mathrm{w}$ is the water fugacity. We assume throughout this work that $r = 1$, which is the nominal value used by \cite{Schaefer:2015} and that expected from experiments on wet diffusion in olivine \citep{Hirth2003}. As in \cite{Schaefer:2015}, we relate the water abundance to fugacity using experimental data on the concentrations of water in olivine from \cite{Li2008} as 
\begin{equation}
\begin{aligned}
\mathrm{ln}f_{\mathrm{w}} & = c_0  + c_1\mathrm{ln}\left(\frac{Bx\mu_{\mathrm{oliv}}/\mu_{\mathrm{w}}}{1-x\mu_{\mathrm{oliv}}/\mu_{\mathrm{w}}}\right) \\ 
& + c_2  \mathrm{ln}^2\left(\frac{Bx\mu_{\mathrm{oliv}}/\mu_{\mathrm{w}}}{1-x\mu_{\mathrm{oliv}}/\mu_{\mathrm{w}}}\right) + c_3 \mathrm{ln}^3 \left(\frac{Bx\mu_{\mathrm{oliv}}/\mu_{\mathrm{w}}}{1-x\mu_{\mathrm{oliv}}/\mu_{\mathrm{w}}}\right)\mathrm{,}
\end{aligned}
\end{equation}
where $c_0 = -7.9859$, $c_1 = 4.3559$, $c_2 = -0.5742$, $c_3 = 0.0227$, $B = 2 \times 10^6$ is a conversion to number concentration (H atom/$10^6$ Si atoms), $\mu_{\mathrm{oliv}}$ is the molecular weight of olivine and $\mu_{\mathrm{w}}$ is the molecular weight of water. As in \cite{Schaefer:2015}, we choose $\eta_0$ such that $\eta(x = x_{\oplus}, T = T_\mathrm{ref}) = 10^{21} \mathrm{Pa} \ \mathrm{s}$ which yields mantle temperatures that approximately reproduce those on Earth. 
\subsubsection{Non-dimensional thermal evolution equation}
\label{sec:nondimtherm}
\begin{table*}
\begin{center}
\begin{tabular}{ c  c  c } 
  \hline 
  \hline
  Quantity & Symbol & Fiducial value \\
  \hline
  Mantle water mass fraction$^{\star}$ & $\twid{x}$ & $\twid{x}_{\oplus} = 1.32$ \\
  Mantle temperature$^{\star}$ & $\twid{T}$ & $\twid{T}_{\mathrm{ref}} = 1$ \\
  Planet mass & $\twid{M}$ & $1$ \\
  Total water mass fraction & $\twid{\omega}$ & 2.07 \\
  Heating timescale & $\tau_{\mathrm{heat}}$ & $t(\tau_{\mathrm{heat}}=1) = 4.02$ Gyr \\
  Pressure-dependent volatile cycling timescale & $\tau$ & $t(\tau=1) = 2.87 $ Gyr \\
  Temperature-dependent volatile cycling timescale & $\tau_{\mathrm{SS}}$ & $t(\tau_{\mathrm{SS}}=1) = 2.22 $ Gyr \\
  Hybrid volatile cycling timescale & $\tau_{\mathrm{hyb}}$ & $t(\tau_{\mathrm{hyb}} = 1) = 2.22$ Gyr \\
  Heat flux  & $ \twid{F}_0$ & 0.531 \\
  Heat flux scaling coefficient & $\beta$ & 0.3 \\
  Critical Rayleigh number & $\mathrm{Ra}_{\mathrm{crit}}$ & 1100 \\
  Water fugacity & $\twid{f}_{\mathrm{w}}$ & 1 \\
  Surface temperature & $\twid{T}_{\mathrm{s}}$ & 0.175 \\
  Reference temperature & $\twid{T}_{\mathrm{m}}$ & 0.040\\
  Mantle water mass fraction of Earth & $\twid{X}_{\oplus}$ & 1.32 \\
  Seafloor pressure degassing exponent & $\mu$ & 1 \\
  Seafloor pressure regassing exponent & $\sigma$ & 1 \\
  Solidus temperature & $\twid{T}_{\mathrm{sol,dry}}$ & 0.780 \\
  Liquidus temperature & $\twid{T}_{\mathrm{liq,dry}}$ & 0.936 \\
  Temperature-dependent degassing coefficient & $\twid{\Pi}$ & 0.102\\
  Solidus depression constant & $\twid{\lambda}$ & $8.16 \times 10^{-5}$ \\
  Solidus depression coefficient & $\gamma$ & 0.75 \\
  Melt fraction exponent & $\theta$ & 1.5 \\
  Pressure-dependent degassing coefficient & $\twid{E}$ & 0.473\\
  Maximum mantle water mass fraction & $\twid{x}_{\mathrm{max}}$ & $15.9$ \\
  \hline
\end{tabular}
\caption {Non-dimensional variables and parameters used in this paper, their symbols, and and their value for Earth-like parameters. Stars denote model state variables.}
\label{table:params}
\end{center}
\end{table*}
Throughout the remainder of this paper, we will work with non-dimensional versions of the thermal evolution and volatile cycling equations. We do so because it elucidates the essential physical processes and controlling non-dimensional variables. Substituting our scaling for mantle heat flux from \Eq{eq:flux} into \Eq{eq:heat} and using our prescription for viscosity from \Eq{eq:viscosity}, we can non-dimensionalize the thermal evolution equation as
\begin{equation}
\label{eq:dTdtnondim}
\begin{aligned}
\frac{d \tilde{T}}{d\tau_{\mathrm{heat}}} = & \ \tilde{Q}(\tau_{\mathrm{heat}}) \\
& - \tilde{F_0}\tilde{f}^{\beta}_{\mathrm{w}}(\tilde{x}) \left(\tilde{T} - \tilde{T_\mathrm{s}}\right)^{\beta + 1} \mathrm{exp}\left[-\frac{\beta}{\tilde{T}_\mathrm{m}}\left(\frac{1}{\tilde{T}} - 1\right)\right] \mathrm{,}
\end{aligned}
\end{equation}
where the non-dimensional temperature is $\tilde{T} = T/T_{\mathrm{ref}}$, the non-dimensional mantle water mass fraction is $\tilde{x} = x f_\mathrm{m}/(\omega_0 \tilde{f}_\mathrm{b}),$ and $\tilde{F_0} = F_0/Q_0$, where $Q_0$ is a constant and 
\begin{equation}
F_0 = \frac{k T^{1+\beta}_{\mathrm{ref}}A}{hV} \left(\frac{\alpha \rho_{\mathrm{m}}g h^3f_{\mathrm{w}}(\tilde{x}=1)}{\mathrm{Ra}_{\mathrm{crit}}\kappa \eta_0}\right)^{\beta}  \mathrm{,}
\end{equation}
where the non-dimensional fugacity is $\tilde{f}_{\mathrm{w}} = f_{\mathrm{w}}/f_{\mathrm{w}}(\tilde{x}=1)$, the reference temperature is $\twid{T}_\mathrm{m} = T_{\mathrm{ref}}R_{\mathrm{gas}}/E_{\mathrm{a}}$, and the surface temperature is $\tilde{T}_\mathrm{s} = T_{\mathrm{s}}/T_{\mathrm{ref}}$. Lastly, the non-dimensional heating timescale is $\tau_{\mathrm{heat}} = t Q_0/(\rho_{\mathrm{m}}c_pT_{\mathrm{ref}})$. The typical values of these non-dimensional parameters are shown in \Tab{table:params}.
\subsection{Volatile cycling}
\indent We seek to explore a variety of different volatile cycling parameterizations, each of which relies on the following expression for the time rate of change of mantle water mass fraction \citep{Cowan2014a}
\begin{equation}
\label{eq:xbas}
\frac{dx}{dt} = \frac{L_{\mathrm{MOR}}S(T)}{f_\mathrm{m}M}\left(w_{\downarrow} - w_{\uparrow}\right) \mathrm{,}
\end{equation}
where $S(T)$ is the temperature-dependent spreading rate (discussed further in \Sec{sec:ss}), $f_\mathrm{m}M$ is the mantle mass (where $f_\mathrm{m}$ is the mantle mass fraction), $w_{\downarrow}$ the regassing rate and $w_{\uparrow}$ the degassing rate. 
Each of the volatile cycling parameterizations we consider utilizes different regassing and degassing rates, which we explore in the following Sections \ref{sec:cowan}-\ref{sec:hybrid}.
\subsubsection{Seafloor pressure-dependent degassing and regassing}
\label{sec:cowan}
In this section, we construct a non-dimensional version of \Eq{eq:xbas} corresponding to the volatile cycling model of \cite{Cowan2014a}. This model determines the water mass fraction of the mantle independent of the mantle temperature. We utilize their expressions for the regassing and degassing rates:
\begin{equation}
\label{eq:wdown}
w_{\downarrow} = x_\mathrm{h} \rho_{\mathrm{c}} d_\mathrm{h}(P) \chi \mathrm{,}
\end{equation}
\begin{equation}
\label{eq:wup}
w_{\uparrow} = x \rho_\mathrm{m}d_{\mathrm{melt}}f_{\mathrm{degas}}(P) \mathrm{,}
\end{equation}
where $x_\mathrm{h}$ is the mass fraction of water in the hydrated crust, $\rho_{\mathrm{c}}$ is the density of the oceanic crust, $\chi$ is the subduction efficiency, $\rho_\mathrm{m}$ is the density of the upper mantle, and $d_{\mathrm{melt}}$ is the depth of melting below mid-ocean ridges. We take $d_{\mathrm{melt}}$ and $\chi$ as constants, with their fiducial value equal to their fiducial value in \cite{Cowan2014a}. As in \cite{Cowan2014a}, we take the depth of the serpentinized layer $d_\mathrm{h}(P)$ and the fraction of the water in the melt that is degassed $f_{\mathrm{degas}}(P)$ to be power-laws with seafloor pressure, with $d_\mathrm{h}$ increasing with increasing pressure and $f_{\mathrm{degas}}$ decreasing with increasing pressure. See \App{sec:cowanderiv} for a thorough explanation of these parameters and the derivation that follows to give the mantle water mass fraction rate of change with time. Inserting Equations (\ref{eq:wdown}) and (\ref{eq:wup}) into \Eq{eq:xbas} and non-dimensionalizing, we find
\begin{equation}
\begin{aligned}
\label{eq:dxdtnondim}
\frac{d\tilde{x}}{d\tau} & = \left[\tilde{g}^2\left(\tilde{\omega} - \tilde{x}\right)\right]^{\sigma} - \tilde{X}^{-1}_{\oplus}\tilde{x} \left[\tilde{g}^2\left(\tilde{\omega} - \tilde{x}\right)\right]^{-\mu} \\
& = \twid{F}_{\downarrow,\mathrm{CA}} - \twid{F}_{\uparrow,\mathrm{CA}}  \mathrm{.}
\end{aligned}
\end{equation}
In \Eq{eq:dxdtnondim}, 
\begin{equation}
\tilde{X}_{\oplus} = \frac{x_h \rho_{\mathrm{c}} \chi d_{\mathrm{h},\oplus} f_\mathrm{M}}{\rho_{\mathrm{m}}d_{\mathrm{melt}} f_{\mathrm{degas},\oplus} \omega_0 \tilde{f}_\mathrm{b}}
\end{equation} 
is a degassing coefficient identified by \cite{Cowan2014a} as the mantle water mass fraction of Earth, the non-dimensional mantle water mass fraction is (as before) $\tilde{x} = x f_\mathrm{m}/(\omega_0 \tilde{f}_\mathrm{b}),$, $\tilde{\omega} = \omega/(\omega_0 \tilde{f}_\mathrm{b})$ is the normalized total water mass fraction, $\twid{g} = g/g_{\oplus}$, and
\begin{equation}
\tau_{\mathrm{CA}} = \tau = t \frac{L_{\mathrm{MOR}}Sx_{\mathrm{h}}\rho_{\mathrm{c}}\chi d_{\mathrm{h},\oplus}}{M \omega_0 \tilde{f}_\mathrm{b}}
\end{equation}
is the non-dimensional time, which is inversely related to the seafloor overturning timescale $A/(L_{\mathrm{MOR}}S)$. The first term on the right hand side of \Eq{eq:dxdtnondim} is the regassing flux $ \twid{F}_{\downarrow,\mathrm{CA}}$ and the second term is the degassing flux $\twid{F}_{\uparrow,\mathrm{CA}}$. In this model, the spreading rate $S$ does not depend on mantle temperature, but it will in Sections \ref{sec:ss} and \ref{sec:hybrid}. We write the non-dimensional timescale here as $\tau$ because it will be the timescale that all of our solutions are converted to for inter-comparison. 
\subsubsection{Temperature-dependent degassing and regassing}
\label{sec:ss}
In this section, we write down a simplified, non-dimensional form of Section 2.3 in \cite{Schaefer:2015}. Their degassing and regassing rates are
\begin{equation}
\label{eq:wdownss}
w_{\downarrow} =  x_\mathrm{h} \rho_\mathrm{c} \chi d_{\mathrm{h}}(T) \mathrm{,}
\end{equation}
\begin{equation}
\label{eq:wupss}
w_{\uparrow} = \rho_{\mathrm{m}} d_{\mathrm{melt}} f_{\mathrm{degas},\oplus} f_{\mathrm{melt}}(T) x  \mathrm{.}
\end{equation}
\Eq{eq:wdownss} is identical to \Eq{eq:wdown} except now the hydrated layer depth is a function of temperature (see \App{sec:ssderiv} for details), and \Eq{eq:wupss} is similar to \Eq{eq:wup} except $f_{\mathrm{degas}}(P)$ has been replaced by $f_{\mathrm{degas},\oplus}f_{\mathrm{melt}}(T)$ with $f_{\mathrm{melt}}(T)$ the temperature-dependent mass fraction. In \Eq{eq:wupss} we have assumed that the mass fraction of water in melt is the same as the mass fraction of water in the mantle due to the extremely low ($\approx 1\%$) difference in water partitioning between melt and mantle rock.  \\
\indent Inserting our expressions (\ref{eq:wdownss}) and (\ref{eq:wupss}) for regassing and degassing rates into \Eq{eq:xbas} and non-dimensionalizing, we find (see \App{sec:ssderiv} for the steps and parameterizations of $S(T),d_{\mathrm{h}}(T),f_{\mathrm{melt}}(T)$)
\begin{equation}
\label{eq:ssvol}
\begin{aligned}
\frac{d\twid{x}}{d\tau_{\mathrm{SS}}} & = \twid{f}^{\beta}_{\mathrm{w}} \left(\twid{T}-\twid{T}_\mathrm{s}\right)^{\beta-1}  \mathrm{exp}\left[\frac{-\beta}{\twid{T}_{\mathrm{m}}}\left(\frac{1}{\twid{T}}-1\right)\right] \\
 & - \twid{\Pi} \twid{f}^{2\beta}_{\mathrm{w}} \left(\twid{T}-\twid{T}_\mathrm{s}\right)^{2\beta}\mathrm{exp}\left[\frac{-2\beta}{\twid{T}_{\mathrm{m}}}\left(\frac{1}{\twid{T}}-1\right)\right]\\
 & \times \twid{x}\left(\twid{T} - \twid{T}_{\mathrm{sol,dry}} + \twid{\lambda} \twid{x}^{\gamma}\right)^{\theta} \\
 & = \twid{F}_{\downarrow,\mathrm{SS}} - \twid{F}_{\uparrow,\mathrm{SS}} \mathrm{.}
\end{aligned}
\end{equation}
In \Eq{eq:ssvol}, the non-dimensional solidus depression coefficient is $\twid{\lambda} = \twid{K}(\omega_0\twid{f}_{\mathrm{b}}/f_{\mathrm{m}})^\gamma$, the degassing coefficient is $\twid{\Pi} = \Pi/D$, where
\begin{equation}
\Pi = \rho_\mathrm{m}d_{\mathrm{melt}}f_{\mathrm{degas},\oplus} \frac{\omega_0 \twid{f}_{\mathrm{b}}}{f_{\mathrm{m}}}\left(\twid{T}_{\mathrm{liq,dry}} - \twid{T}_{\mathrm{sol,dry}}\right)^{-\theta}  \mathrm{,}
\end{equation}
and the regassing coefficient (related to the hydrated layer depth) is
\begin{equation}
D = x_\mathrm{h}\rho_{\mathrm{c}}\chi_{\mathrm{r}}h^{(1-3\beta)}\frac{(T_{\mathrm{serp}} - T_{\mathrm{s}})}{T^{1+\beta}_{\mathrm{ref}}} \left(\frac{\kappa \mathrm{Ra}_{\mathrm{crit}} \eta_0}{\alpha\rho_{\mathrm{m}}gf_{\mathrm{w}}(\twid{x}=1)}\right)^{\beta} 	\mathrm{,}
\end{equation}
and $\tau_{\mathrm{SS}} = t D/\Sigma$, where 
\begin{equation}
\Sigma = M\omega_0 \twid{f}_{\mathrm{b}}\frac{h^{(1-6\beta)}\kappa^{(2\beta-1)}}{10.76L_{\mathrm{MOR}}}\left(\frac{\mathrm{Ra}_{\mathrm{crit}}\eta_0}{\alpha\rho_{\mathrm{m}}gT_{\mathrm{ref}}f_{\mathrm{w}}(\twid{x}=1)}\right)^{2\beta} 	
\end{equation}
is related to the spreading rate. Additonally, we have re-expressed the first term on the right hand side of \Eq{eq:ssvol} as the regassing flux $ \twid{F}_{\downarrow,\mathrm{SS}}$ and the second term on the right hand side as the degassing flux $\twid{F}_{\uparrow,\mathrm{SS}}$. \\
\indent In our coupled integrations of Equations (\ref{eq:dTdtnondim}) and (\ref{eq:ssvol}) we ensure that the hydrated layer does not contain more water than the surface in order to maintain water mass balance \citep{Schaefer:2015}. In terms of our analytic model, this is equivalent to ensuring that the regassing coefficient $D$ (which is related inversely to the non-dimensional degassing coefficient $\tilde{\Pi}$) never exceeds a critical value, which is written in \Eq{eq:d2constraint}. As a result, this is a constraint on the rate of subduction of water that ensures that the amount of water in the mantle never exceeds the total amount of water in the planet. 
\subsubsection{Seafloor pressure-dependent degassing and temperature-dependent regassing}
\label{sec:hybrid}
In this section, we construct a model where the degassing rate is determined by seafloor pressure (as volcanism rates will be lower if overburden pressure is higher) and the regassing rate is determined by the mantle temperature (as the depth of serpentinization will be lower if temperature is higher). 
We construct such a model because serpentinization can only happen below a critical temperature, whereas it has not been conclusively shown to depend on overburden pressure. Meanwhile, it has been shown that volcanism rates on exoplanets should be inversely related to the overburden pressure \citep{Kite2009a}. In this model, the degassing rate is taken from \Eq{eq:wup} with $\mu=1$ (the value expected from \citealp{Kite2009a}) and the regassing rate from \Eq{eq:wupss}. Using the same method as in Sections \ref{sec:cowan} and \ref{sec:ss}, we substitute these into \Eq{eq:xbas} and non-dimensionalize (see \App{sec:hybderiv} for more details). Doing so, we find 
\begin{equation}
\label{eq:hybridx}
\begin{aligned}
\frac{d\twid{x}}{d\tau_{\mathrm{hyb}}} & = \twid{f}^{\beta}_{\mathrm{w}} \left(\twid{T}-\twid{T}_\mathrm{s}\right)^{(\beta-1)} \mathrm{exp}\left[\frac{-\beta}{\twid{T}_\mathrm{m}}\left(\frac{1}{\twid{T}} - 1\right)\right] \\
& - \twid{E} \twid{f}^{2\beta}_{\mathrm{w}} \left(\twid{T}-\twid{T}_\mathrm{s}\right)^{(2\beta)} \mathrm{exp}\left[\frac{-2\beta}{\twid{T}_\mathrm{m}}\left(\frac{1}{\twid{T}} - 1\right)\right] \\
& \times \twid{x} \left[\twid{g}^2 \left(\twid{\omega}-\twid{x}\right)\right]^{-1} \\
& = \twid{F}_{\downarrow,\mathrm{hyb}} - \twid{F}_{\uparrow,\mathrm{hyb}}\mathrm{,}
\end{aligned}
\end{equation}
where $\twid{E} = E/D$, $E = \rho_\mathrm{m}d_{\mathrm{melt}}f_{\mathrm{degas},\oplus}\omega_0\twid{f}_\mathrm{b}/f_{\mathrm{m}}$, and $\tau_{\mathrm{hyb}} = \tau_{\mathrm{SS}} = tD/\Sigma$. As before, we have re-written the first term on the right hand side as the regassing flux $\twid{F}_{\downarrow,\mathrm{hyb}}$ and the second term on the right hand side as the degassing flux $\twid{F}_{\uparrow,\mathrm{hyb}}$.
\section{Comparison of volatile cycling parameterizations}
\subsection{Time-dependent}
\label{sec:comparison}
\begin{figure*}
	\centering
	\includegraphics[width=.85\textwidth]{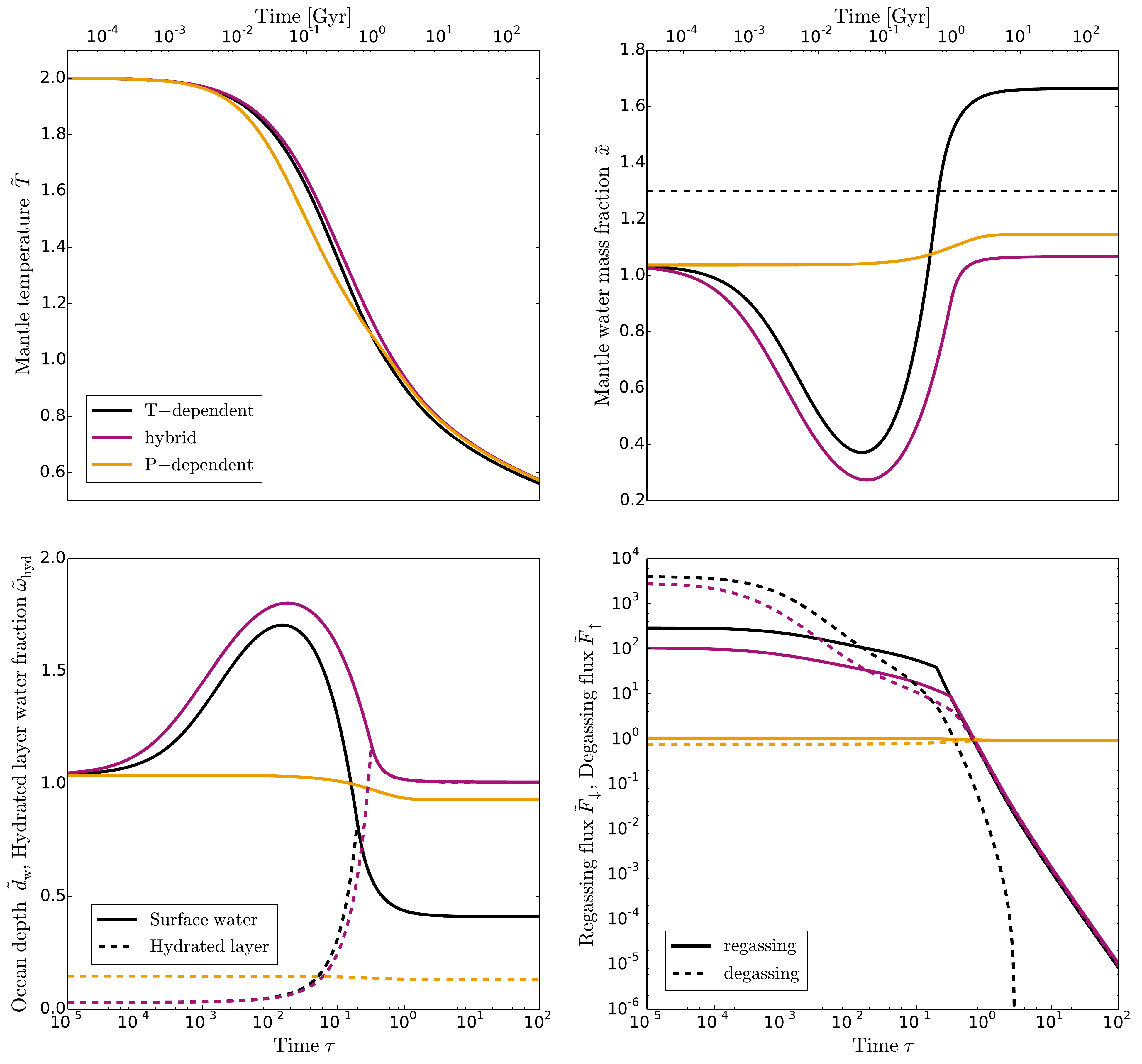}
	\caption{Comparison between evolution of temperature, mantle water mass fraction, ocean depth (in this case equivalent to surface water mass fraction), hydrated layer water mass fraction, and regassing and degassing fluxes for models in Sections \ref{sec:cowan}, \ref{sec:ss}, and \ref{sec:hybrid}. Integrations were performed for Earth-like parameter values: $\twid{M} = 1$, $\twid{\omega} = 2.07$, $\twid{x}_0 = \twid{\omega}/2$, $\twid{T}_0 = 2$. The dashed line on the mantle water mass fraction plot shows the estimated present-day Earth value. Note that because here we use an ocean basin covering fraction $\twid{f}_{\mathrm{b}}$ of $1.3$ times that of Earth, this is not a direct analogue of Earth. The ocean depth at which the model would result in a waterworld is $\twid{d}_{\mathrm{w}} = 2.85$, far above the maximum found in all three models. All models reach an eventual steady-state (or near steady-state in the case of the temperature-dependent model) in mantle water mass fraction, although their mantles perpetually cool. We show evolution well past the age of the Solar System to display the stability of the steady-states achieved. ``P-dependent'' corresponds to the model in \Sec{sec:cowan}, ``T-dependent'' the model in \Sec{sec:ss}, and ``hybrid'' the model in \Sec{sec:hybrid}.}
	\label{fig:comparison}
\end{figure*}
\indent Before turning to the steady-state solutions, we compare directly the time-dependent evolution of the three models. \Fig{fig:comparison} shows such a comparison for Earth-like parameters. The hybrid and pressure-dependent models reach a steady state after a time $\tau \approx 1$, with a value that is independent of initial conditions (not shown). Meanwhile, the solely temperature-dependent model reaches a near steady-state where a tiny amount of net regassing still occurs. \Fig{fig:comparison} shows that although the mantle temperature evolution does not vary by more than $\sim 15\%$ among models, the evolution and steady-state value of mantle water mass fraction varies greatly. Notably, the degassing parameterizations lead to different values of the end-state mantle water mass fraction $\twid{x}$ even though both the hybrid and solely temperature-dependent models have their late volatile evolution determined by water mass balance between the hydrated layer and surface. We will explain this in detail in the steady-state solutions of \Sec{sec:steadystatesol}.  \\
\indent \Fig{fig:comparison} also shows the evolution of the non-dimensional ocean depth for each of the models considered. To determine the ocean depth for a given $\twid{x}$, we utilize Equation (15) of \cite{Cowan2014a}. This relates ocean depth to seafloor pressure by $\twid{d}_{\mathrm{w}} = P/(g\rho_{\mathrm{w}}$), where $d_\mathrm{w}$ is the ocean depth and $\rho_\mathrm{w}$ the density of water. Non-dimensionalizing, we find 
\begin{equation}
\label{eq:dw}
\twid{d}_{\mathrm{w}} = \twid{g}\left(\twid{\omega} - \twid{x}\right) \mathrm{,}
\end{equation}
where $\twid{d}_{\mathrm{w}} = d_{\mathrm{w}}/d_{\mathrm{w},\oplus}$ is the ocean depth normalized to that of Earth, $d_{\mathrm{w},\oplus} = 4$ km. Note that gravity comes into \Eq{eq:dw} from the ratio of planetary mass to area, which is proportional to gravity. Because we use an Earth-like $\twid{g} = 1$ in these calculations, the ocean depth is equivalent to the surface water mass fraction $\twid{\omega}-\twid{x}$. The evolution of ocean depth with time hence has an opposite sign to that for the mantle water mass fraction, with the hybrid model having the deepest oceans and the \cite{Schaefer:2015} model the least end-state surface water. \\
\indent Alongside the ocean depth (equivalently surface water mass fraction) we show the evolution of the mass fraction of water in the hydrated layer $\twid{\omega}_{\mathrm{hyd}} = M_{\mathrm{hyd}}/(\omega_0\twid{f}_{\mathrm{b}}M)$, where $M_{\mathrm{hyd}}$ is the mass of water in the hydrated layer. We can relate the mass fraction of water in the hydrated layer and the hydrated layer depth as
\begin{equation}
\twid{\omega}_{\mathrm{hyd}} = \frac{4\pi}{3\omega_0\twid{f}_{\mathrm{b}}}\left[R^3 - (R-d_\mathrm{h})^3\right]\frac{x_\mathrm{h}\rho_{\mathrm{m}}}{M} \mathrm{,}
\end{equation}  
where as in \cite{Cowan2014a} we use $x_{\mathrm{h}} = 0.05$ and $\rho_{\mathrm{m}} = 3.3 \times 10^{3} \ \mathrm{kg} \ \mathrm{m}^{-3}$. For the pressure-dependent model, we calculate the hydrated layer thickness from \Eq{eq:dhp} using $d_{\mathrm{h},\oplus} = 3 \ \mathrm{km}$. For the temperature-dependent and hybrid models, we calculate the hydrated layer thickness using \Eq{eq:dhydr} unless the hydrated layer thickness limit is violated, in which case we calculate it from Equations (\ref{eq:dhydr2}) and (\ref{eq:d2constraint}). As shown in \Fig{fig:comparison}, the amount of water in the hydrated layer increases drastically in the temperature-dependent and hybrid models when regassing begins to dominate over degassing. 
At late times in the temperature-dependent and hybrid models, the hydrated layer water mass fraction is equal to the surface mass fraction. As discussed further in \Sec{sec:compEarth}, this is not representative of present-day Earth. However, the hydrated layer water mass fraction in the pressure-dependent model stays small at all times, as in this model the hydrated layer thickness is a simple power-law with pressure. \\
\indent \Fig{fig:comparison} also shows the individual regassing and degassing rates (the first and second terms on the right hand side of Equations (\ref{eq:dxdtnondim}), (\ref{eq:ssvol}), and (\ref{eq:hybridx})). The models in \Sec{sec:ss} and \Sec{sec:hybrid} have an initial phase of degassing from the mantle followed by strong regassing of water back to the mantle. Note that this initial phase of degassing occurs over a shorter timescale than the low-viscosity (``boundary-layer'') model of \cite{Schaefer:2015}, as we assume that all of the water is in melt (thereby increasing the amount able to be outgassed) and use a higher initial mantle temperature. We can qualitatively understand the varied evolution in different models by examining how the degassing and regassing rates vary with temperature and/or pressure. Note that the initial phase of degassing from the mantle occurs whether the degassing is temperature or seafloor pressure-dependent, as long as regassing of water back into the mantle is temperature-dependent. This is because initially the regassing rate is very low due to the smaller hydrated layer thickness when the mantle is hot. Regassing then becomes more efficient as the mantle temperature drops and the hydrated layer thickness grows. Similarly, degassing becomes less efficient at later times. This is because the mantle is cooler and hence has a lower melt fraction and the seafloor pressure is greater, both decreasing the rate of volcanism. \\
\indent However, the end-state evolution for these models is slightly different. For the \cite{Schaefer:2015} model, the degassing rate drops to zero at late times because of the lack of melt available for degassing, while the regassing rate (though small) is non-negligible. As a result, there is net regassing at late times in our temperature-dependent model. The hybrid model, meanwhile, does reach a true steady-state. This is because the degassing rate stays large at late times, as it is not dependent on the mantle convection itself, and because the hydrated layer thickness limit is reached, which sharply decreases the regassing rate. Meanwhile, the regassing rate is limited by the hydrated layer thickness, which reaches the limit given by \Eq{eq:d2constraint} after $\sim 1$ Gyr of evolution. This regassing rate then decreases more strongly with time due to the constancy of the maximum hydrated layer thickness and the decreasing spreading rate with decreasing mantle temperature, leading to convergence of the degassing and regassing rates and entrance into steady-state. Such a steady-state was not found in the volatile cycling models of \cite{Sandu2011} and \cite{Schaefer:2015}, which utilized parameterized convection. This is because the evolution was either too slow to reach steady-state over the age of the observable universe or because slow ingassing continued to deplete the surface water reservoir. The former occurs in models that consider the viscosity as an average mantle viscosity rather than that relevant for the interface between the boundary layer and mantle interior. We do the latter in this work. As discussed in \Sec{sec:deepintro}, Earth is likely currently at or near a steady-state in surface water mass fraction. We can estimate the steady-state ocean depths for our models as a function of planetary parameters, and will do so in \Sec{sec:steadystatesol}. 
\subsection{Comparison to Earth}
\label{sec:compEarth}
\begin{table*}
\begin{center}
\begin{tabular}{ c  c  c  c  c } 
  \hline 
  \hline
   & Pressure-dependent & Temperature-dependent & Hybrid & Earth  \\
  \hline
  Mantle water mass fraction $\twid{x}$ & 1.14 (0.050\%) & 1.65 (0.072\%) & 1.06 (0.047\%) & 1.3 (0.057\%)  \\
  Surface water mass fraction $\twid{\omega} - \twid{x}$ & 0.931 (0.028\%) & 0.424 (0.013\%) & 1.01 (0.030\%) & 0.75 (0.022\%) \\
  Upper mantle temperature $\twid{T}$ & 0.866 (1386 K) & 0.844  (1350 K)& 0.875 (1400 K) & $0.8 - 1.2$ (1280 - 1920 K) \\
   Ocean depth $\twid{d}_\mathrm{w}$ & 0.931 (3.72 km) & 0.424 (1.70 km) & 1.01 (4.04 km) & 0.75 (3.0 km) \\
  Hydrated layer mass fraction $\twid{\omega}_{\mathrm{hyd}}$ & 0.132 ($3.9 \times 10^{-3}$\%) & 0.423 (0.013\%) & 1.01 (0.030\%) & 0.14 ($4.23 \times 10^{-3}$\%) \\
  Degassing flux ${\twid{F}_{\uparrow}}$ & 0.928 & $2.55 \times 10^{-3}$  & 0.105 & 0.03  \\
  & ($6.06 \times 10^{12}$ kg yr$^{-1}$) & ($1.67 \times 10^{10}$ kg yr$^{-1}$) & ($6.86 \times 10^{11}$ kg yr$^{-1}$) & ($2 \times 10^{11}$ kg yr$^{-1}$) \\
  Regassing flux ${\twid{F}_{\downarrow}}$ & 0.932 & 0.0973 & 0.119 & $0.1 - 0.4$  \\
  & ($6.09 \times 10^{12}$ kg yr$^{-1}$) & ($6.36 \times 10^{11}$ kg yr$^{-1}$) & ($7.78 \times 10^{11}$ kg yr$^{-1}$) & ($[0.7-2.9] \times 10^{12}$ kg yr$^{-1}$) \\
  \hline
\end{tabular}
\caption{Values of key model variables (mantle water mass fraction, surface water mass fraction, upper mantle temperature, ocean depth, mass fraction of water in the hydrated layer, degassing and regassing fluxes) after $4.5 \ \mathrm{Gyr}$ of evolution for the Earth-like case shown in \Fig{fig:comparison}. Approximate present-day Earth values from \cite{Hirschmann2006} and \cite{Cowan2014a} are shown for reference. Dimensionful values are shown in parentheses. The value of Earth's ocean depth is computed given an ocean basin covering fraction ($\twid{f}_{\mathrm{b}}$) of $1.3$, which is larger that for current Earth and leads to a smaller ocean depth than seen.  Note that Earth's mantle water mass fraction and regassing and degassing rates are very approximate, accurate at best to within a factor of $\sim 2$ \citep{Cowan2014a}. The uncertainty in characteristic Earth upper mantle temperatures stems from the range of possible relevant depths.}
\label{table:comp}
\end{center}
\end{table*}
\indent Given that each of the models considered produces different mantle and volatile evolution, here we compare their results to that of Earth. We do so in order to understand which model(s) may be most physical for application to exoplanets. We compare the key variables (mantle water mass fraction, surface water mass fraction, upper mantle temperature, ocean depth, hydrated layer mass fraction, degassing and regassing rates) derived from our Earth-like model after $4.5$ Gyr of evolution to those of Earth itself in \Tab{table:comp}. Note that we are using an ocean basin covering fraction that is $1.3$ times that of Earth, so we calculate equivalent ocean depths for this increased ocean basin area. \\ 
\indent No model matches well Earth's present-day water partitioning between mantle and surface, with the estimated Earth mantle water mass fraction lying between those of the temperature-dependent and pressure-dependent models. As mentioned previously, the temperature-dependent and hybrid models well over-predict the amount of water in the hydrated layer. However, the pressure-dependent model matches well the estimated mass fraction of water in the hydrated layer. Though there is considerable uncertainty in Earth's mantle water mass fraction and regassing and degassing rates, the hybrid model has reasonably similar values to both of these. The pressure-dependent model over-predicts both the degassing and regassing fluxes, as these do not decrease strongly with time in this model. However, note that the degassing and regassing fluxes estimated from observations for Earth do not match, so if Earth water cycling is in steady-state one of these must be erroneous by approximately an order of magnitude. The temperature-dependent model reasonably matches Earth's present-day regassing rate, but well over-predicts the mantle water mass fraction and under-predicts the degassing rate, due to the lack of degassing at late times. \\
\indent In general, none of the models alone match all of the Earth constraints, though each model does approximate at least one constraint. Given that the temperature-dependent and hybrid models have almost all of their surface water in the hydrated layer, the pressure-dependent model is most representative of present-day Earth. The pressure-dependent model is also closest to the present-day surface water mass fraction of Earth. Though it matches Earth's regassing flux within a factor of two, it over-predicts the degassing flux. However, if the degassing and regassing fluxes of present-day Earth are in steady-state, including a necessary increase in degassing flux such that it matches the regassing flux would allow the pressure-dependent model to match all available constraints. If not, future work is needed to develop a model that matches well all of the available constraints from Earth.
\subsection{Steady-state mantle water mass fraction}
\label{sec:steadystatesol}
Given that all of our models reach a steady-state in water partitioning on the timescale of a few billion years, we examine steady-state solutions to the models in Sections \ref{sec:cowan}-\ref{sec:hybrid}. We do so because these steady-states are the most observationally relevant, as most planets in the habitable zone will lie around $\sim \mathrm{Gyr}$-age main-sequence stars.
We note that due to continuous regassing, the steady-state for the \cite{Schaefer:2015} model is one where the amount of surface water is simply determined by the amount of water that can be incorporated into the mantle of a planet. This is the ``petrological limit'' of the mantle, and will be discussed in detail in \Sec{sec:waterworld} below. Note that if the total water mass fraction is less than the petrological limit, the mantle holds all of the water except that which remains on the surface due to mass-balance with the hydrated layer. Solving for the steady-state of the pressure-dependent and hybrid models using Equations (\ref{eq:dxdtnondim}) and (\ref{eq:hybridx}) gives
\begin{equation}
\label{eq:sstates1}
\twid{x} = \tilde{X}_{\oplus} \left[\twid{g}^2(\twid{\omega}-\twid{x})\right]^{\mu+\sigma}
\end{equation}
for the \cite{Cowan2014a} model in \Sec{sec:cowan} and 
\begin{equation}
\label{eq:sstates3}
\twid{x} = \twid{\omega} \left(1 + \twid{E}\twid{f}^{\beta}_{\mathrm{w}}\left(\twid{T}-\twid{T}_\mathrm{s}\right)^{1+\beta} \mathrm{exp}\left[\frac{-\beta}{\twid{T}_{\mathrm{m}}}\left(\frac{1}{\twid{T}}-1\right)\right]\twid{g}^{-2}\right)^{-1}
\end{equation}
for the hybrid model in \Sec{sec:hybrid}. We choose $\twid{T} = \twid{T}_{\mathrm{sol,dry}}$ to calculate the steady-state mantle water mass fraction for the hybrid model. This is because the steady-state is nearly independent of temperature for $\twid{T} \lesssim \twid{T}_{\mathrm{m}}$ (see \Fig{fig:comparison}).
\Eq{eq:sstates1} reproduces Equation (20) of \cite{Cowan2014a}. Note that the steady-state value of $\twid{x}$ for the model in \Sec{sec:ss} is independent of $\twid{\omega}$. This greatly limits the relative amount of water that can be put into the mantle of planets with large total water fractions. \\
\indent Equations (\ref{eq:sstates1}) and (\ref{eq:sstates3}) give us transcendental expressions for the steady-state mantle water mass fraction as a function of mantle temperature and planet mass for each model. In \Sec{sec:waterworld} we solve Equations (\ref{eq:sstates1}) and (\ref{eq:sstates3}) and relate the mantle water mass fraction to the surface ocean depth to determine the waterworld limit for various assumptions about the processes that control volatile cycling on exoplanets. 
\section{What determines if a planet will be a waterworld?}
\label{sec:waterworld}
\begin{figure}
	\centering
	\includegraphics[width=.5\textwidth]{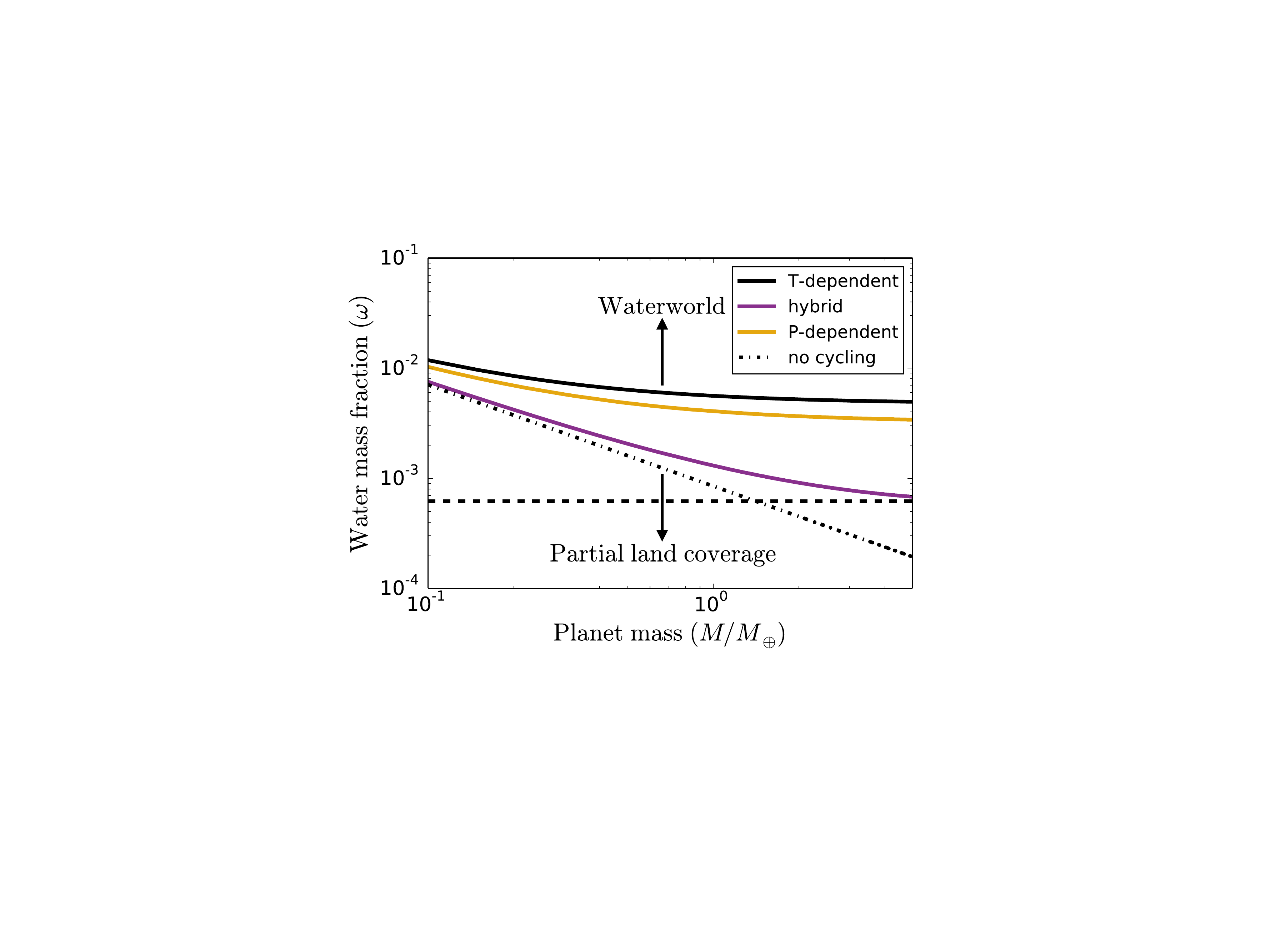}
	\caption{Waterworld boundary as a function of total water mass fraction and planet mass (normalized to that of Earth) for the volatile cycling models considered. All fixed parameters are at their fiducial values (\Tab{table:params}). Planets above each line are waterworlds, and planets below the line have partial land coverage. The dashed line shows the approximate value of Earth's total water mass fraction assuming that the mantle has $2.5$ Earth ocean masses of water \citep{Cowan2014a}. The dot-dashed line shows what the waterworld boundary would be if water cycling did not occur and all of the planetary water resided on the surface. The pressure-dependent model of \cite{Cowan2014a} predicts that planets require a much larger total mass of water to become waterworlds than the hybrid model, but has a similar waterworld boundary to that of the temperature-dependent model from \cite{Schaefer:2015}. The minimum water mass fraction to become a waterworld for the hybrid model decreases strongly with planet mass, meaning that super-Earths are more likely to become waterworlds if degassing is temperature-independent but regassing temperature-dependent.}
	\label{fig:wwbound}
\end{figure}
In this section, we use our steady-state solutions from \Sec{sec:steadystatesol} to make predictions of the minimum total water mass fraction needed to become a waterworld for a given planet mass. As in \cite{Cowan2014a}, in this calculation we keep $\twid{x}$ below its petrological limit of $0.7\%$ by mass or $12$ ocean masses for an Earth-mass planet with a perovskite mantle. In our notation, this means $\twid{x}_{\mathrm{max}} = 15.9$. The mantles of super-Earths will be largely post-perovskite \citep{Valencia2007a}, which may hold more water than perovskite, up to $\approx 2\%$ by mass \citep{Townsend2015}. We do not include such a phase transition in our model, but note that an increase in the maximum water fraction linearly translates to an increase in the total water fraction at which a planet becomes a waterworld. This will be explored further in the sensitivity analysis of \Sec{sec:sensitivity}. \\
 \indent To determine whether a planet is a waterworld, we compare the steady-state ocean depth calculated from \Eq{eq:dw} to the maximum depth of water-filled ocean basins (Equation 11 of \citealp{Cowan2014a})
\begin{equation}
\twid{d}_{\mathrm{o,max}} \approx \frac{d_{\mathrm{o,max},\oplus}}{d_{\mathrm{w},\oplus}}\twid{g}^{-1} \mathrm{,}
\end{equation}
where $d_{\mathrm{o,max},\oplus} = 11.4$ km. This maximum depth comes from isostatic arguments which consider the maximum thickness that continents can achieve before they flow under their own weight, and adopting the crustal thickness of the Himalayan plateau ($70 \ \mathrm{km}$) as this limit. If $\twid{d}_{\mathrm{w}} > \twid{d}_{\mathrm{o,max}}$, the planet is a waterworld. 
\\ \indent \Fig{fig:wwbound} shows the waterworld boundary as a function of total water mass fraction and planet mass for all three volatile cycling models considered here. We show the predictions up to $5$ Earth masses, as this is near where the transition between rocky and gaseous exoplanets lies \citep{Lopez2014,Rogers2015}. As shown in \cite{Cowan2014a}, mantle temperature-independent volatile cycling models predict that a large water mass fraction ($0.3 - 1\%$) is needed for a planet to become a waterworld, with only a slight dependence on planet mass. The model of \cite{Schaefer:2015} predicts a similar but slightly larger water fraction than that of \cite{Cowan2014a}. This is because the mantle in the \cite{Schaefer:2015} model is at the petrologic limit of water mass fraction. The model of \cite{Cowan2014a} is near this limit, and as we show in \Sec{sec:sensitivity} hits the limit if the seafloor pressure dependence $\sigma + \mu$ is increased by $50\%$ from its nominal value.  \\
\indent The hybrid model, meanwhile, predicts that a much lower total water mass fraction is needed for a planet to become a waterworld. The limiting water mass fraction decreases more strongly with increasing planet mass in this model, meaning that super-Earths are more likely to be waterworlds if the hybrid model is physically relevant. However, this limiting water mass fraction remains larger than in the case without volatile cycling (dot-dashed line in \Fig{fig:wwbound}). Notably, the waterworld boundary reaches Earth's water mass fraction for $\approx 5 M_{\oplus}$ planets. This is because the hybrid model does not have temperature-dependent degassing, and therefore degassing does not decrease strongly in efficacy at late times when the mantle is cool. Instead, degassing of water at mid-ocean ridges reaches a true steady state with the temperature-dependent regassing when the surface complement of water becomes deep enough to slow down the degassing rate and the regassing rate is limited by the maximum depth of the hydrated layer. This is unlike the \cite{Schaefer:2015} model, in which a near steady-state is only reached because there is a limit to the rate and amount of total regassing set through the maximum hydrated layer thickness. Instead, it is more similar to weakening the pressure-dependence of the \cite{Cowan2014a} model from power-law exponents $\sigma + \mu = 2$ (their nominal model) to $\sigma + \mu = 1$, as the hybrid model effectively sets their degassing exponent $\mu = 1$ and regassing exponent $\sigma = 0$.  As we discuss in \Sec{sec:observations}, the stark differences between the waterworld water-mass limit in the hybrid model and the \cite{Cowan2014a} and \cite{Schaefer:2015} may be potentially observable.   
\section{Discussion}
\label{sec:discussion}
\subsection{Sensitivity Analysis}
\label{sec:sensitivity}
\begin{figure}
	\centering
	\includegraphics[width=.5\textwidth]{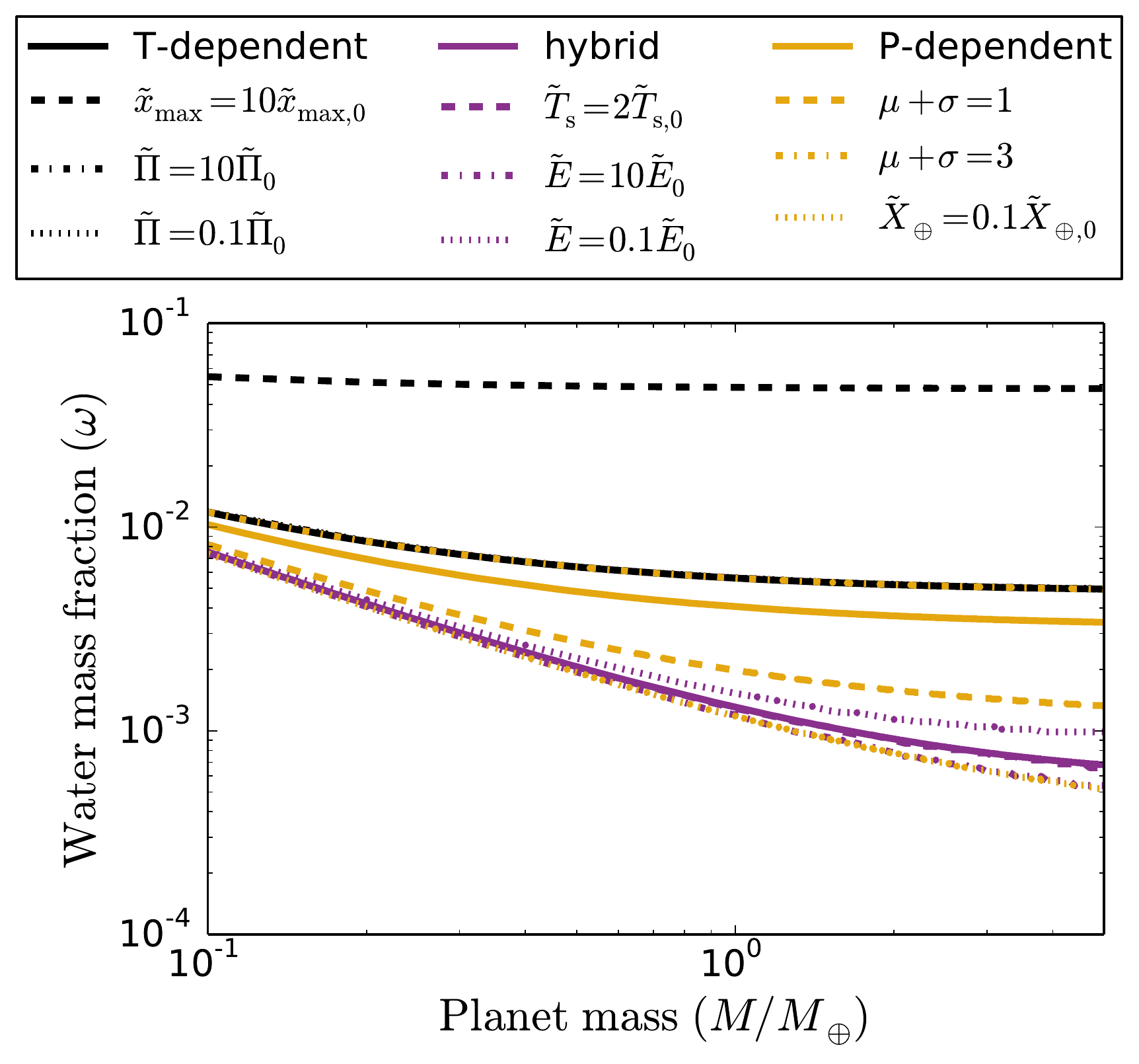}
	\caption{Analysis of the sensitivity of the waterworld boundary to varying non-dimensional parameters. The solid lines reproduce the waterworld boundary from \Fig{fig:wwbound}, while dashed, dashed-dotted, and dotted lines with a given color show the changes in the wateworld boundary for the corresponding model. For the \cite{Schaefer:2015} model, increasing the maximum value of mantle water mass fraction correspondingly moves the waterworld boundary up in total water mass fraction. Varying $\twid{\Pi}$ has no effect because the model has already increased the mantle water mass fraction to its maximal value. For the \cite{Cowan2014a} model, decreasing $\mu + \sigma$ decreases the total water fraction needed to become a waterworld. Increasing $\mu + \sigma$ increases the waterworld boundary, but the mantle reaches its petrological limit of mantle water mass fraction if $\mu + \sigma = 3$. Decreasing $\twid{X}_{\oplus}$ decreases the waterworld boundary by a similar fraction, and increasing $\twid{X}_{\oplus}$ similarly increases the boundary until the maximum water mass fraction is reached (not shown). For the hybrid model, varying surface temperature plays little role in changing the waterworld boundary. Increasing $\twid{E}$ slightly decreases the water mass fraction to become a waterworld, but this is a relatively small effect. Our conclusion that the total water mass fraction needed to become a waterworld is much smaller for the hybrid model is hence robust to uncertainties in parameter values.}
	\label{fig:wwbound_paramstudy}
\end{figure}
\indent In this section, we perform a sensitivity analysis to determine how the non-dimensional parameters affect our steady-state solutions from \Sec{sec:steadystatesol}. The key unknown parameters that affect our solutions are the maximum mantle water mass fraction $\twid{x}_{\mathrm{max}}$ (which affects all models), Earth mantle water mass fraction $\twid{X}_{\oplus}$ and seafloor pressure power-law exponents $\mu + \sigma$ for the model of \cite{Cowan2014a}, degassing coefficient $\twid{\Pi}$ for the \cite{Schaefer:2015} model, degassing coefficient $\twid{E}$ in the hybrid model, and surface temperature $\twid{T}_{\mathrm{s}}$ for both the \cite{Schaefer:2015} and hybrid models. Importantly, our steady-states are independent of the abundance of radiogenic elements, eliminating some of the natural variation between planetary systems. Though the abundance of radiogenic elements affects the time it takes to reach steady-state, the steady-state volatile cycling is independent of the decreasing mantle temperature at late times. \Fig{fig:wwbound_paramstudy} shows how varying these parameters in each model affects our derived waterworld boundary. \\
\indent Though the changes in the waterworld boundary with changing $\mu + \sigma$ have been explored in \cite{Cowan2014a}, we reproduce them here for comparison with the other models. Decreasing the dependencies of degassing and regassing on seafloor pressure reduces the water mass fraction at which the surface is completely water-covered, with a maximum decrease of a factor of $2$ between the $\mu + \sigma = 2$ and $\mu + \sigma = 1$ cases. Similarly, increasing the dependence to $\mu + \sigma = 3$ increases the limiting water mass fraction to become a  waterworld, but the model reaches the maximum mantle water mass fraction. If $\twid{X}_{\oplus}$ is a factor of ten lower than used here, the waterworld boundary decreases by a comparably large fraction, especially for super-Earths. If $\twid{X}_{\oplus}$ is much larger than assumed here, the mantle will be at its petrological limit of water intake and the waterworld boundary will be determined by the maximum mantle water mass fraction. \\
\indent For the model of \cite{Schaefer:2015}, which is at the petrological limit of maximum mantle water mass fraction, varying $\twid{\Pi}$ by an order of magnitude in either direction does not change the waterworld boundary. However, increasing the maximum mantle water mass fraction by a given value increases the total water mass fraction needed to become a waterworld by a comparable amount. For both the hybrid model and the \cite{Schaefer:2015} model, changing the surface temperature only leads to minute changes in the waterworld boundary. This is because the surface temperature cannot vary by more than a factor of a few or else liquid water would not be stable on the surface. Increasing $\twid{E}$ in the hybrid model decreases the total mass fraction needed to become a waterworld, but by less than a factor of two for all masses. 
\begin{figure}
	\centering
	\includegraphics[width=.5\textwidth]{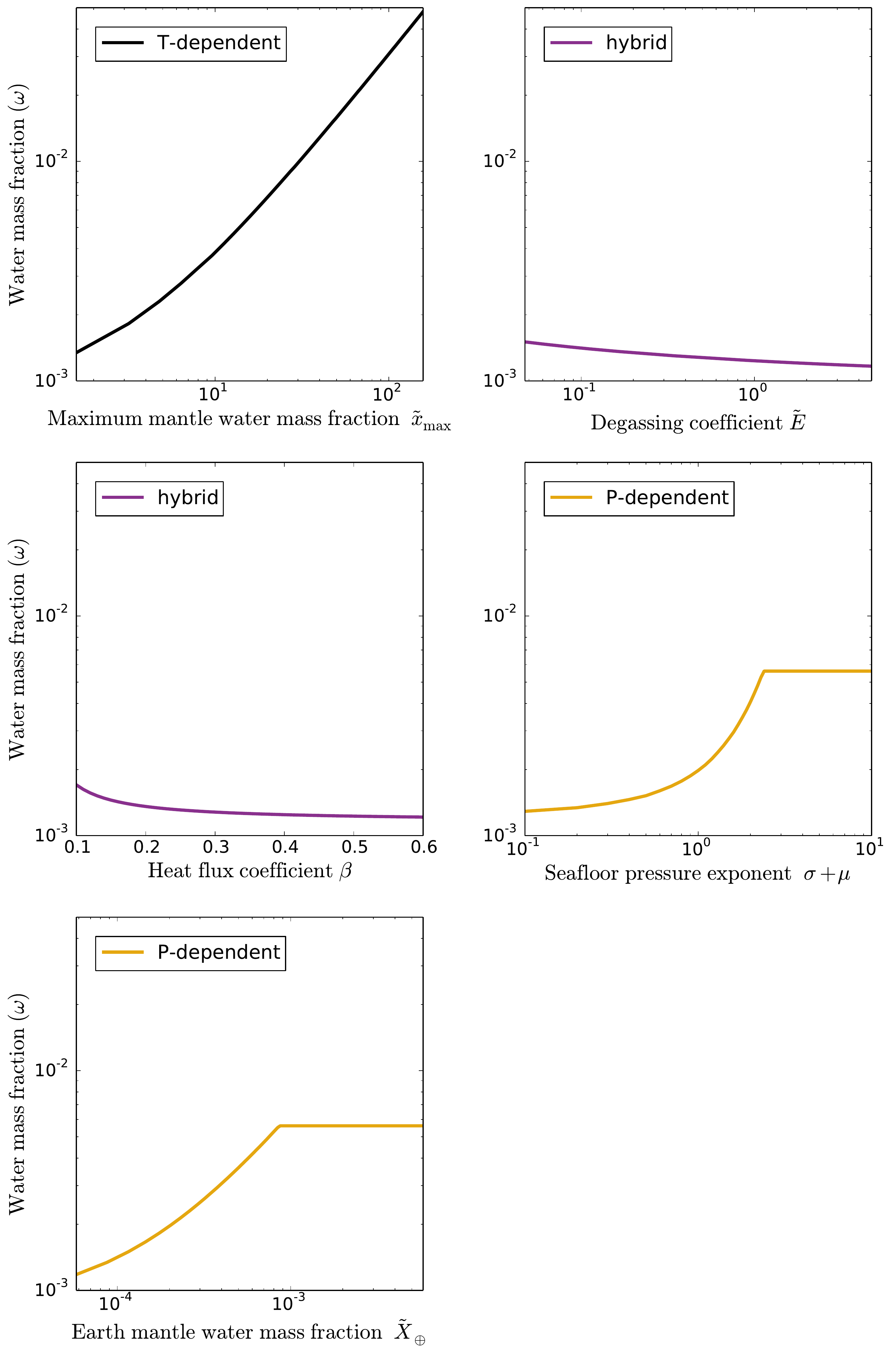}
	\caption{Sensitivity analysis on the waterworld boundary to varying non-dimensional parameters with a fixed $M/M_{\oplus} = 1$. We vary the parameters that have the largest impact on the waterworld boundary: $\twid{x}_{\mathrm{max}}$ for the \cite{Schaefer:2015} model, $\twid{E}$ and $\beta$ for the hybrid model, and $\sigma + \mu$ and $\twid{X}_{\oplus}$ for the \cite{Cowan2014a} model. The waterworld boundary for the \cite{Schaefer:2015} model is strongly dependent on $\twid{x}_{\mathrm{max}}$. The boundary for the \cite{Cowan2014a} model is dependent on $\sigma + \mu$ and $\twid{X}_{\oplus}$ up to the limit where the mantle becomes saturated with water. The results of the hybrid model are only marginally sensitive to $\twid{E}$ and $\beta$, giving us confidence that the hybrid model indeed lowers the planetary water mass fraction needed to become a waterworld.}
	\label{fig:wwbound_paramstudy_constantmass}
\end{figure}
\\ \indent From \Fig{fig:wwbound_paramstudy}, we can identify four key non-dimensional parameters that change the waterworld boundary by a sizeable amount: $\twid{x}_{\mathrm{max}}$ (which mainly affects the \citealp{Schaefer:2015} model), $\twid{E}$ for the hybrid model, and $\sigma + \mu$ and $\twid{X}_{\oplus}$ for the \cite{Cowan2014a} model. \Fig{fig:wwbound_paramstudy_constantmass} shows how continuously varying these parameters by one order of magnitude around their fiducial value with a planet mass fixed equal to that of Earth affects the water mass fraction at which planets become waterworlds. We also consider varying $\beta$, which could affect the solution since a reduced outgoing flux would lead to larger mantle temperatures and hence larger steady-state mantle water mass fractions. As mentioned above, our results are very sensitive to the petrological limit of the mantle water mass fraction, but an order of magnitude increase in $\twid{E}$, $\sigma + \mu$, and $\twid{X}_{\oplus}$ leads to only a factor of $\sim 2$ or less increase in the waterworld limit. The waterworld boundary is also largely insensitive to $\beta$, which should not vary by more than a factor of $2$ from its nominal value of $0.3$. Note that increasing both $\sigma + \mu$ and $\twid{X}_{\oplus}$ cannot lead to continuous increases in the waterworld limit, as the petrological limit of water mantle mass fraction is reached just above our fiducial values for these parameters. As a result, the \cite{Cowan2014a} model is, like the \cite{Schaefer:2015} model, sensitive to the maximum mantle water mass fraction $\twid{x}_{\mathrm{max}}$. \\
\indent Our results are much less sensitive to $\twid{E}$, for which an order-of-magnitude increase only decreases the waterworld boundary by $\sim 10\%$. Note that the non-dimensional degassing rate $\twid{E}$ scales with the normalization of mantle viscosity as $\twid{E} \propto \eta^{-\beta}_0$. As a result, if the viscosities were increased (for instance, in the case that the middle-mantle viscosity is more relevant for the convection parameterization), the degassing rate would decrease as a power-law with increasing viscosity. However, because our solutions are only weakly dependent on $\twid{E}$, the choice of mantle viscosity does not greatly affect the waterworld boundary itself. We explore the effects of a larger viscosity further in \Sec{sec:scaling}, as it will substantially affect the evolutionary timescales for water cycling. In general, the conclusion that super-Earths are more likely to be waterworlds if degassing is temperature-independent is robust to order-of-magnitude uncertainties in our non-dimensional parameters. 
\subsection{Scaling of timescales with planet mass and mantle viscosity}
\label{sec:scaling}
\indent Though the waterworld boundary itself is largely robust to varying our non-dimensional parameters, the timescale to reach steady-state depends on the planet mass and mantle viscosity. In this section, we derive how the evolution timescale varies with these parameters in order to determine the mass regime at which planets may not reach steady-state. The evolution timescale scales with planet mass as $t(\tau=1) \propto M^{1-p}$, if $S$ is independent of mass (as it is in the pressure-dependent model). Similarly,  $t(\tau=1) \propto M^{1 - 2\beta p - 2\beta}$ if $S$ is dependent on mass (as it is in the temperature-dependent and hybrid models, see Equation (\ref{eq:sratefin})) where $p = 0.27$ and $\beta \approx 0.3$. As a result, the evolution of mantle water fraction is slower for larger planets and is only weakly sensitive to mass, with $t(\tau=1) \propto M^{0.24}$, in the more realistic case where $S$ depends on planet mass. \\
\indent The evolution timescale scales with the viscosity as $t(\tau_{\mathrm{SS}}=1) \propto \eta^{\beta}_0$. As a result, the evolution timescale is also a power-law in viscosity, increasing with increasing viscosity. The choice of a characteristic mantle viscosity hence may affect the resulting mantle evolution, with an order of magnitude increase in viscosity leading to a factor of $\sim 2$ increase in the timescale to reach steady-state. \\ 
\indent Based on the scaling of the evolutionary timescales with mass alone, our conclusion that water cycling reaches steady-state is unaffected. This is because a $5M_{\oplus}$ planet would only take $\approx 1.5$ times longer to reach steady-state than Earth. However, if the viscosity normalization is more than a factor of $\approx 5$ larger than assumed here, the evolution would take longer than the age of the Solar System to reach steady state. This could occur if the viscosity of the deep mantle is relevant for our parameterized convection scheme, and is similar to the conclusion from the high-viscosity models of \cite{Schaefer:2015}. However, these models were not shown to produce water cycling evolution similar to Earth, while we showed in \Sec{sec:compEarth} that a boundary-layer viscosity can match some of the constraints from Earth. In general, it is clear that understanding in detail which characteristic viscosity is relevant for parameterized convection is necessary to make more robust predictions of water cycling on exoplanets.
\subsection{Comparison with previous work}
\indent In this work, we developed simplified models for water cycling between the mantle and the surface based on previous models in the literature. We did so in order to compare their predictions for whether or not terrestrial exoplanets will be waterworlds. This is the first such test of physical assumptions that has been performed for volatile cycling on planets with varying mass, though \cite{Sandu2011} explored how models of varying complexity affect volatile cycling on Earth. Our models find that the surface water mass fraction reaches a steady-state after $\sim 2$ Gyr of evolution. Though this has been found when keeping the ratio of the degassing and regassing rates fixed in time \citep{McGovern1989}, no steady-state has previously been found when these rates are dependent on the mantle temperature and allowed to separately evolve. \\
\indent We find a steady-state in our models for two reasons: the degassing rate is initially larger than the regassing rate (leading to convergent evolution of the two rates), and the volatile evolution is relatively quick. In our temperature-dependent and hybrid models, the volatile evolution is quick because we use a viscosity relevant for the upper mantle, leading to faster evolution than for a viscosity relevant in the deep mantle \citep{Schaefer:2015}. The use of an upper mantle viscosity here is reasonable since it is physically motivated from boundary-layer theory (see \Sec{sec:convbackground}) and better matches Earth's current near-steady state. Additionally, the regassing rate is limited because the mass of water in the hydrated layer cannot be greater than that on the surface, which as in \cite{Schaefer:2015} leads to a sharp decrease in the regassing rate at late times. Given that the degassing rate is also small due to the low temperatures, these rates balance to determine our steady-states and hence waterworld boundary limits from \Sec{sec:waterworld}. \\
\indent We agree with the conclusion of \cite{Cowan2014a} that super-Earths are unlikely to be waterworlds for both of the models with solely seafloor pressure-dependent and temperature-dependent water cycling. We also similarly find that conclusion is likely unaffected by parameter uncertainties. However, if these models themselves are less physical than a model with seafloor pressure dependencies dominating the degassing rate and temperature dependencies controlling regassing, that conclusion may change. In this hybrid model, the total mass fraction needed to become a waterworld is much smaller than that in both the \cite{Cowan2014a} and \cite{Schaefer:2015} models. Ideally, future work will help distinguish between the viability of the three models considered here.
\subsection{Limitations}
\indent In this work we considered three separate parameterizations for water cycling between ocean and mantle in order to make predictions for how they might affect exoplanet surface water abundances. We did so because the processes that control volatile cycling on Earth are not understood to the level of detail needed to make predictions for exoplanets with varying masses, total water mass fractions, compositions, and climates. Due to this, we utilized a simplified semi-analytic model and parameterized volatile cycling rates as either a power-law in pressure or a function of temperature. In general, though this simplified model is powerful for understanding how a given process changes the surface water budget of the suite of exoplanets, studying surface water evolution on a given planet enables the use of more detailed coupling of parameterized convection and volatile cycling as in \cite{Sandu2011,Schaefer:2015}. \\
\indent There also remain important parameters that do not have well-characterized dependencies with planet mass. Similarly to \cite{Schaefer:2015}, we identified that the characteristic mantle viscosity is an important unknown in the problem, as it can affect the evolutionary timescales. Additionally, the maximum mantle water mass fraction alone determines the waterworld boundary for the temperature-dependent model, and it is not known exactly how this should depend on planet mass. Understanding these parameters in detail will be necessary to make more detailed predictions of volatile cycling on terrestrial exoplanets.
\subsection{Observational constraints and future work}
\label{sec:observations}
\indent It is clear that there is a dichotomy in the waterworld boundary based on whether or not one assumes that volatile cycling is temperature-dependent and/or pressure-dependent. As a result, understanding better which processes control volatile cycling on Earth is important to make more stringent predictions of whether terrestrial exoplanets should be waterworlds. Alternatively, observations with post-\textit{JWST}-era instruments may be able to determine whether or not there is exposed land through either infrared spectra (if the atmosphere is not too optically thick, \citealp{Abbot2012}) or photometric observations over an entire planetary orbit in many wavelengths \citep{Cowan2009,Kawahara2010,Cowan2013,Cowan2014a}. This would serve as a test of the different volatile cycling parameterizations. If some super-Earths are found to have non-zero land fraction, the hybrid model considered here is not important or volatile delivery is inefficient for these objects. If, on the other hand, super-Earths are found to all be waterworlds, considering the combined effects of seafloor pressure-limited degassing and mantle convection may be necessary to explain volatile cycling on terrestrial planets. 
 \\
\indent In the future, one could use sophisticated multi-dimensional calculations of mantle convection including degassing through mid-ocean ridge volcanism and regassing through subduction of hydrated basalt, but this would be computationally expensive. However, these sophisticated calculations will not be worthwhile until the specific processes that govern volatile cycling on Earth and terrestrial exoplanets are understood in detail. We propose, then, that future observations of terrestrial exoplanets will be able to distinguish between the various water cycling models considered in this work. This could help constrain theories for water cycling on Earth and enable more sophisticated models to make predictions for the surface water inventory of individual planets. However, we must first understand in detail the effects of early water delivery and loss, and the effects of various tectonic regimes on water cycling itself.
\section{Conclusions}
\label{sec:conc}
\begin{enumerate}
\item Volatile cycling on terrestrial exoplanets with plate tectonics should reach an approximate steady-state on the timescale of a few billion years, independent of the volatile cycling parameterization used. Given that Earth is likely near a steady-state in surface water mass fraction, this gives us confidence that many terrestrial exoplanets around main-sequence stars are also at or near steady-state. The steady states in the temperature-dependent and hybrid models may be substantially different from present-day Earth, as both these
models store approximately an order of magnitude more water in the hydrated crust than Earth itself.
\item Models considering either temperature-dependent degassing and regassing or pressure-dependent degassing and regassing predict that copious amounts of water ($\sim 0.3-1\%$ of total planetary mass) must be present to form a waterworld. These models have their mantles saturated with water, and if the total water mass fraction is high they are at or near the petrological limit for how much water the mantle can hold. The waterworld boundary for the solely temperature-dependent volatile cycling model is determined by this limit. As a result, if a super-Earth mantle can hold more water, the waterworld boundary will move upward by a similar factor. This would make it even less likely for super-Earths to be waterworlds.
\item If seafloor pressure is important for the degassing rate of water but not for regassing, it is more likely that super-Earths will be waterworlds. In this case, a super-Earth with the same total water mass fraction as Earth could become a waterworld. These planets would be less likely to be habitable, as unlucky planets with a large amount of initial water delivery may lack a silicate weathering feedback to stabilize their climates. Understanding further which processes determine volatile cycling on Earth will help us understand what processes control mid-ocean ridge degassing and subduction rates of water on exoplanets with surface oceans. 
\end{enumerate}

\acknowledgements
This work was aided greatly by discussions with L. Coogan, N. Cowan, C. Goldblatt, A. Lenardic, L. Schaefer, N. Sleep, and K. Zahnle. We thank the anonymous referee for helpful comments that greatly improved the manuscript. We thank the Kavli Summer Program in Astrophysics for the setting to perform this research and the hospitality of the program members and community at the University of California, Santa Cruz. TDK acknowledges support from NASA headquarters under the NASA Earth and Space Science Fellowship Program Grant PLANET14F-0038. DSA acknowledges support from the NASA Astrobiology Institute Virtual Planetary Laboratory, which is supported by NASA under cooperative agreement NNH05ZDA001C. 

\appendix
\section{Volatile cycling schemes: derivation}
\label{app:deriv}
\subsection{Seafloor-pressure dependent degassing and regassing} 
\label{sec:cowanderiv}
In this section, we write down a time-dependent version of the model from \cite{Cowan2014a}, where degassing and regassing are regulated by seafloor pressure. The regassing and degassing rates in this case are
\begin{equation}
\label{eq:wdown2}
w_{\downarrow} = x_\mathrm{h} \rho_{\mathrm{c}} d_\mathrm{h}(P) \chi \mathrm{,}
\end{equation}
\begin{equation}
\label{eq:wup2}
w_{\uparrow} = x \rho_\mathrm{m}d_{\mathrm{melt}}f_{\mathrm{degas}}(P) \mathrm{,}
\end{equation}
equivalent to Equations  (\ref{eq:wdown}) and (\ref{eq:wup}). Here $x_\mathrm{h}$ is the mass fraction of water in the hydrated crust, $\rho_{\mathrm{c}}$ the density of the oceanic crust, $\chi$ the subduction efficiency, $\rho_\mathrm{m}$ the density of the upper mantle, $d_{\mathrm{melt}}$ the depth of melting below mid-ocean ridges, $d_{\mathrm{h}}$ the hydrated layer depth and $f_{\mathrm{degas}}$ the degassing efficiency. To derive \Eq{eq:dxdtnondim}, we start with \Eq{eq:xbas} and substitute Equations (\ref{eq:wdown2}) and (\ref{eq:wup2})
\begin{equation}
\label{eq:dxdt2}
\frac{dx}{dt} = \frac{L_{\mathrm{MOR}}S}{f_\mathrm{M}M}\bigg[x_\mathrm{h} \rho_{\mathrm{c}} \chi d_{\mathrm{h},\oplus}\left(\frac{P}{P_{\oplus}}\right)^\sigma - x\rho_\mathrm{m}d_{\mathrm{melt}} f_{\mathrm{degas},\oplus}\left(\frac{P}{P_{\oplus}}\right)^{-\mu}\bigg] \mathrm{,}
\end{equation}
where $L_{\mathrm{MOR}} = 3\pi R_{\mathrm{p}}$ is the mid-ocean ridge length, $S$ is the average spreading rate of Earth ($\approx 10 \ \mathrm{cm} \ \mathrm{year}^{-1}$), and we have used the power laws
\begin{equation}
\label{eq:dhp}
d_\mathrm{h}(P) = d_{\mathrm{h},\oplus}\left(\frac{P}{P_{\oplus}}\right)^\sigma \mathrm{,}
\end{equation}
\begin{equation}
\label{eq:fdegas}
f_{\mathrm{degas}}(P) = f_{\mathrm{degas},\oplus}\left(\frac{P}{P_{\oplus}}\right)^{-\mu} \mathrm{.}
\end{equation}
\cite{Cowan2014a} chose power-laws to illustrate how different strengths of seafloor pressure-dependence would operate. In \Eq{eq:dhp} $d_{\mathrm{h},\oplus}$ is the hydration depth on Earth, $P_{\oplus}$ is Earth's seafloor pressure, and in \Eq{eq:fdegas} $f_{\mathrm{degas},\oplus}$ is the melt degassing fraction on modern Earth. Note that seafloor pressure $P = g \rho_\mathrm{w} d_\mathrm{w}$, where $d_\mathrm{w}$ is the ocean depth and $\rho_\mathrm{w}$ the density of water. \\
\indent \cite{Cowan2014a} relate seafloor pressure to mantle water mass fraction by
\begin{equation}
P = P_{\oplus} \tilde{g}^2\frac{(\omega - x f_\mathrm{M})}{\omega_0\tilde{f_\mathrm{b}}} \mathrm{,}
\end{equation}
where $\omega_0 = 2.3 \ \times 10^{-4}$ is the fractional mass of Earth's surface water and $\tilde{f_{\mathrm{b}}} = f_b/f_{b,\oplus} = 1.3$ is the ocean basin covering fraction normalized to that of Earth. Plugging this expression for $P$ into \Eq{eq:dxdt2}, we find
\begin{equation}
\label{eq:dxdtfin}
\frac{dx}{dt} = \frac{L_{\mathrm{MOR}}S}{f_\mathrm{M}M}\bigg[x_\mathrm{h} \rho_{\mathrm{c}} \chi d_{\mathrm{h},\oplus}\left(\tilde{g}^2\frac{(\omega - x f_\mathrm{M})}{\omega_0\tilde{f_\mathrm{b}}}\right)^\sigma - \rho_\mathrm{m}d_{\mathrm{melt}} x f_{\mathrm{degas},\oplus}\left(\tilde{g}^2\frac{(\omega - x f_\mathrm{M})}{\omega_0\tilde{f_\mathrm{b}}}\right)^{-\mu}\bigg] \mathrm{.} \\
\end{equation}
Note that we can write \Eq{eq:dxdtfin} using $\twid{\omega} = \omega/(\omega_0\twid{f}_{\mathrm{b}})$ and $\twid{x} = xf_{\mathrm{m}}/(\omega_0\twid{f}_{\mathrm{b}})$ as
\begin{equation}
\label{eq:dxdtfin_alt}
\frac{dx}{dt} = \frac{L_{\mathrm{MOR}}S}{f_\mathrm{M}M}\bigg[x_\mathrm{h} \rho_{\mathrm{c}} \chi d_{\mathrm{h},\oplus}\left[\tilde{g}^2(\twid{\omega} - \twid{x})\right]^\sigma - \rho_\mathrm{m}d_{\mathrm{melt}} x f_{\mathrm{degas},\oplus}\left[\tilde{g}^2(\twid{\omega} - \twid{x})\right]^{-\mu}\bigg] \mathrm{.}
\end{equation}
Non-dimensionalization of \Eq{eq:dxdtfin_alt} then gives 
\begin{equation}
\label{eq:dxdtnondim2}
\frac{d\tilde{x}}{d\tau} = \left[\tilde{g}^2\left(\tilde{\omega} - \tilde{x}\right)\right]^{\sigma} - \tilde{X}^{-1}_{\oplus}\tilde{x} \left[\tilde{g}^2\left(\tilde{\omega} - \tilde{x}\right)\right]^{-\mu} \mathrm{,}
\end{equation}
equivalent to \Eq{eq:dxdtnondim}. In \Eq{eq:dxdtnondim2}, 
\begin{equation}
\tilde{X}_{\oplus} = \frac{x_h \rho_{\mathrm{c}} \chi d_{\mathrm{h},\oplus} f_\mathrm{M}}{\rho_{\mathrm{m}}d_{\mathrm{melt}} f_{\mathrm{degas},\oplus} \omega_0 \tilde{f}_\mathrm{b}}
\end{equation} 
is the non-dimensionalized mantle water mass fraction of Earth, $\tilde{\omega} = \omega/(\omega_0 \tilde{f}_\mathrm{b})$ is the non-dimensionalized total water mass fraction, $\twid{g} = g/g_{\oplus}$, and
\begin{equation}
\tau_{\mathrm{CA}} = \tau = t \frac{L_{\mathrm{MOR}}Sx_{\mathrm{h}}\rho_{\mathrm{c}}\chi d_{\mathrm{h},\oplus}}{M \omega_0 \tilde{f}_\mathrm{b}}
\end{equation}
is the non-dimensional time, which is inversely related to the seafloor overturning timescale $A/(L_{\mathrm{MOR}}S)$.
\subsection{Temperature-dependent degassing and regassing}
\label{sec:ssderiv}
In this section, we derive a simplified version of the \cite{Schaefer:2015} model, where volatile cycling rates are determined by the mantle temperature. The regassing and degassing rates in this case are
\begin{equation}
\label{eq:wdownss2}
w_{\downarrow} =  x_\mathrm{h} \rho_\mathrm{c} \chi d_{\mathrm{h}}(T) \mathrm{,}
\end{equation}
\begin{equation}
\label{eq:wupss2}
w_{\uparrow} = \rho_{\mathrm{m}} d_{\mathrm{melt}} f_{\mathrm{degas},\oplus} f_{\mathrm{melt}}(T) x  \mathrm{,}
\end{equation}
equivalent to  Equations (\ref{eq:wdownss}) and (\ref{eq:wupss}). Here we have written the hydrated layer depth as a function of temperature. We have written $f_{\mathrm{degas}}$ as $f_{\mathrm{degas},\oplus}f_{\mathrm{melt}}(T)$ where $f_{\mathrm{melt}}(T)$ is the temperature-dependent melt fraction. Inserting Equations (\ref{eq:wdownss2}) and (\ref{eq:wupss2}) into \Eq{eq:xbas}, the dimensionful time-derivative of mantle water mass fraction is 
\begin{equation}
\label{eq:SSwaterfull}
\frac{dx}{dt} = \frac{L_{\mathrm{MOR}}S(T)}{f_\mathrm{m}M}\left[x_\mathrm{h}\rho_{\mathrm{c}}\chi_\mathrm{r}d_{\mathrm{h}}(T) - \rho_\mathrm{m}d_{\mathrm{melt}}f_{\mathrm{degas},\oplus}f_{\mathrm{melt}}(T)x \right] \mathrm{.}
\end{equation}
The functional forms of $S,d_{\mathrm{h}},f_{\mathrm{melt}}$ are developed in Section 2.3 of \cite{Schaefer:2015}. Here we simplify them in order to obtain an analytically tractable version of \Eq{eq:SSwaterfull}. Firstly, the spreading rate is defined as
\begin{equation}
\label{eq:srate}
S = 2u_{\mathrm{conv}} = 2 \frac{5.38 \kappa h}{\delta^2} \mathrm{,}
\end{equation}
The boundary-layer thickness $\delta$ is
\begin{equation}
\label{eq:delta2}
\delta = h \left(\frac{\mathrm{Ra}_{\mathrm{crit}}}{\mathrm{Ra}}\right)^{\beta} \mathrm{.}
\end{equation}
Substituting $\delta$ from \Eq{eq:delta2},
\begin{equation}
\label{eq:sratefin}
S  = \frac{10.76 \kappa}{h} \left(\frac{\mathrm{Ra}}{\mathrm{Ra}_{\mathrm{crit}}}\right)^{(2\beta)} = 10.76 \kappa^{(1-2\beta)}h^{(6\beta - 1)} \left(\frac{\alpha\rho_{\mathrm{m}}g(T-T_{\mathrm{s}})}{\eta(T,x) \mathrm{Ra}_{\mathrm{crit}}}\right)^{2\beta} \mathrm{.}
\end{equation}
\indent The hydration depth (depth to which rock can be serpentinized) is defined as
\begin{equation}
d_{\mathrm{h}} = k \frac{(T_{\mathrm{serp}} - T_{\mathrm{s}})}{F_{\mathrm{m}}} \mathrm{.}
\end{equation}
Using the mantle heat flux from \Eq{eq:flux}, we find 
\begin{equation}
\label{eq:dhydr}
d_{\mathrm{h}} = h \frac{(T_{\mathrm{serp}} - T_{\mathrm{s}})}{(T - T_\mathrm{s})}\left(\frac{\mathrm{Ra}_{\mathrm{crit}}}{\mathrm{Ra}}\right)^\beta = h^{(1-3\beta)} (T-T_\mathrm{s})^{-(1+\beta)}(T_{\mathrm{serp}}-T_\mathrm{s})\left(\frac{\eta(T,x)\kappa \mathrm{Ra}_{\mathrm{crit}}}{\alpha\rho_{\mathrm{m}}g}\right)^{\beta} \mathrm{.}
\end{equation}
\indent Lastly, we use the same expression for the melt fraction as \cite{Schaefer:2015}, which relates the melt fraction to mantle temperature through a power-law, taking into account the solidus depression of wet mantle 
\begin{equation}
\label{eq:fmelt}
f_{\mathrm{melt}} = \left(\frac{T - T_{\mathrm{sol,wet}}(x)}{T_{\mathrm{liq,dry}}-T_{\mathrm{sol,dry}}}\right)^{\theta} \mathrm{.}
\end{equation}
Here, we take $T_{\mathrm{liq,dry}} \approx 1498 \ \mathrm{K}$, $T_{\mathrm{sol,dry}} \approx 1248 \ \mathrm{K}$ as constants, and $ T_{\mathrm{sol,wet}} = T_{\mathrm{sol,dry}} - Kx^{\gamma}$, assuming that the mass fraction of water in melt is the same as the mass fraction of water in the mantle. We assume so because the partitioning coefficient of water in the mantle is thought to be extremely small ($\approx 1\%$). Plugging in Equations (\ref{eq:sratefin}, \ref{eq:dhydr}, \ref{eq:fmelt}) into \Eq{eq:SSwaterfull} and non-dimensionalizing gives
\begin{equation}
\label{eq:ssvol2}
\begin{aligned}
\frac{d\twid{x}}{d\tau_{\mathrm{SS}}} = &  \twid{f}^{\beta}_{\mathrm{w}} \left(\twid{T}-\twid{T}_\mathrm{s}\right)^{\beta-1}  \mathrm{exp}\left[\frac{-\beta}{\twid{T}_{\mathrm{m}}}\left(\frac{1}{\twid{T}}-1\right)\right] \\
& - \twid{\Pi} \twid{f}^{2\beta}_{\mathrm{w}} \left(\twid{T}-\twid{T}_\mathrm{s}\right)^{2\beta}\mathrm{exp}\left[\frac{-2\beta}{\twid{T}_{\mathrm{m}}}\left(\frac{1}{\twid{T}}-1\right)\right] \twid{x}\left(\twid{T} - \twid{T}_{\mathrm{sol,dry}} + \twid{\lambda} \twid{x}^{\gamma}\right)^{\theta} \mathrm{,}
\end{aligned}
\end{equation}
equivalent to \Eq{eq:ssvol}. Here the non-dimensional solidus depression coefficient is $\twid{\lambda} = \twid{K}(\omega_0\twid{f}_{\mathrm{b}}/f_{\mathrm{m}})^\gamma$ and the degassing coefficient is $\twid{\Pi} = \Pi/D$, where
\begin{equation}
\Pi = \rho_\mathrm{m}d_{\mathrm{melt}}f_{\mathrm{degas},\oplus} \frac{\omega_0 \twid{f}_{\mathrm{b}}}{f_{\mathrm{m}}}\left(\twid{T}_{\mathrm{liq,dry}} - \twid{T}_{\mathrm{sol,dry}}\right)^{-\theta}  \mathrm{.}
\end{equation}
The regassing coefficient (related to the hydrated layer depth) is
\begin{equation}
D = x_\mathrm{h}\rho_{\mathrm{c}}\chi_{\mathrm{r}}h^{(1-3\beta)}\frac{(T_{\mathrm{serp}} - T_{\mathrm{s}})}{T^{1+\beta}_{\mathrm{ref}}} \left(\frac{\kappa \mathrm{Ra}_{\mathrm{crit}} \eta_0}{\alpha\rho_{\mathrm{m}}gf_{\mathrm{w}}(\twid{x}=1)}\right)^{\beta} \mathrm{,}
\end{equation}
and $\tau_{\mathrm{SS}} = t D/\Sigma$, where 
\begin{equation}
\Sigma = M\omega_0 \twid{f}_{\mathrm{b}}\frac{h^{(1-6\beta)}\kappa^{(2\beta-1)}}{10.76L_{\mathrm{MOR}}}\left(\frac{\mathrm{Ra}_{\mathrm{crit}}\eta_0}{\alpha\rho_{\mathrm{m}}gT_{\mathrm{ref}}f_{\mathrm{w}}(\twid{x}=1)}\right)^{2\beta} \mathrm{.}
\end{equation} \\
\indent To ensure water mass balance in their  time-dependent solutions, \cite{Schaefer:2015} force the hydrated layer to hold no more water than the surface itself. Formally, this ensures that
\begin{equation}
x_h \rho_\mathrm{m}\frac{4\pi}{3}\left(R^3 - (R-d_{\mathrm{h}})^3\right) \le M\omega_0\twid{f}_\mathrm{b}\left(\twid{\omega} - \twid{x}\right) \mathrm{.}
\end{equation} 
Noting that we can re-write the hydrated layer depth from \Eq{eq:dhydr} as
\begin{equation}
\label{eq:dhydr2}
d_{\mathrm{h}} = D_2 \mathrm{exp}\left[\frac{\beta}{\twid{T}_\mathrm{m}}\left(\frac{1}{\twid{T}} - 1\right)\right] \twid{f}^{-\beta}_{\mathrm{w}}\left(\twid{T}-\twid{T}_\mathrm{s}\right)^{-(1+\beta)} \mathrm{,}
\end{equation}
where $D_2 = D/(x_h\rho_\mathrm{c}\chi_\mathrm{r})$, we find a constraint for $D_2$ to ensure that the hydrated layer water mass is less than or equal to that on the surface:
\begin{equation}
\label{eq:d2constraint}
D_2 \le \left[R - \left(R^3 - \frac{3\omega_0 \twid{f}_\mathrm{b} M(\twid{\omega}-\twid{x})}{4 \pi x_\mathrm{h} \rho_{\mathrm{m}}}\right)^{1/3}\right] \mathrm{exp}\left[\frac{-\beta}{\twid{T}_\mathrm{m}}\left(\frac{1}{\twid{T}} - 1\right)\right] \twid{f}^{\beta}_{\mathrm{w}}\left(\twid{T}-\twid{T}_\mathrm{s}\right)^{(1+\beta)} \mathrm{.}
\end{equation}
We force the constraint from \Eq{eq:d2constraint} in each timestep to ensure stability\footnote{If this constraint is not placed, the mantle water mass fraction will go to infinity.}. Using the maximum value of $D_2$, we can find the maximum value of $\twid{\Pi}$ for use in \Eq{eq:ssvol}
\begin{equation}
\label{eq:pimax}
\twid{\Pi}_{\mathrm{max}} = \frac{\Pi}{D_{2,\mathrm{max}}x_h\rho_\mathrm{c}\chi_\mathrm{r}} \mathrm{.} 
\end{equation}
\subsection{Seafloor pressure-dependent degassing and temperature-dependent regassing}
\label{sec:hybderiv}
Given the above models with either temperature or seafloor pressure-dependent volatile cycling rates, one can envision a model where surface water abundance is regulated by both seafloor pressure and mantle temperature. Here we consider a hybrid model where seafloor pressure regulates the degassing rate (as volcanism is less efficient with greater overburden pressure) and mantle temperature regulates the regassing rate (because serpentinization cannot occur if temperatures are too high). This hybrid model follows similarly from our derivations in \App{sec:cowanderiv} and \App{sec:ssderiv}. The regassing and degassing rates in this case are
\begin{equation}
\label{eq:wdown3}
w_{\downarrow} = x_\mathrm{h} \rho_\mathrm{c} \chi d_{\mathrm{h}}(T) \mathrm{,}
\end{equation}
\begin{equation}
\label{eq:wupss3}
w_{\uparrow} = x \rho_\mathrm{m}d_{\mathrm{melt}}f_{\mathrm{degas}}(P) \mathrm{,}
\end{equation}
equivalent to Equations (\ref{eq:wdown2}) and (\ref{eq:wupss2}). Inserting these into \Eq{eq:xbas}, we find the dimensional form of the time-derivative of water mass fraction
\begin{equation}
\label{eq:dxdthyb}
\frac{dx}{dt} = \frac{S(T)}{f_\mathrm{m}M}\left[x_\mathrm{h}\rho_{\mathrm{c}}\chi_\mathrm{r}d_{\mathrm{h}}(T) - \rho_\mathrm{m}d_{\mathrm{melt}}x f_{\mathrm{degas}}(P)\right] \mathrm{.}
\end{equation}
We insert our prescriptions for $S$ and $d_{\mathrm{h}}$ from Equations (\ref{eq:sratefin}) and (\ref{eq:dhydr}), respectively, and the seafloor pressure-dependence of $f_{\mathrm{degas}}$ from \Eq{eq:fdegas} into \Eq{eq:dxdthyb}. Non-dimensionalizing, we find
\begin{equation}
\label{eq:hybridx2}
\frac{d\twid{x}}{d\tau_{\mathrm{hyb}}} = \twid{f}^{\beta}_{\mathrm{w}} \left(\twid{T}-\twid{T}_\mathrm{s}\right)^{(\beta-1)} \mathrm{exp}\left[\frac{-\beta}{\twid{T}_\mathrm{m}}\left(\frac{1}{\twid{T}} - 1\right)\right] - \twid{E} \twid{f}^{2\beta}_{\mathrm{w}} \left(\twid{T}-\twid{T}_\mathrm{s}\right)^{(2\beta)} \mathrm{exp}\left[\frac{-2\beta}{\twid{T}_\mathrm{m}}\left(\frac{1}{\twid{T}} - 1\right)\right] \twid{x} \left[\twid{g}^2 \left(\twid{\omega}-\twid{x}\right)\right]^{-1} \mathrm{,}
\end{equation}
equivalent to \Eq{eq:hybridx}. Here $\twid{E} = E/D$, $E = \rho_\mathrm{m}d_{\mathrm{melt}}f_{\mathrm{degas},\oplus}\omega_0\twid{f}_\mathrm{b}/f_{\mathrm{m}}$, and $\tau_{\mathrm{hyb}} = \tau_{\mathrm{SS}} = tD/\Sigma$.  \\
\indent As in to the solely temperature-dependent model, we restrict the hydrated layer depth using \Eq{eq:d2constraint}. If $D_2 = D_{2,\mathrm{max}}$ the corresponding constraint on $\twid{E}$ is
\begin{equation}
\label{eq:Emax}
\twid{E}_{\mathrm{max}} = \frac{E}{D_{2,\mathrm{max}}x_h\rho_\mathrm{c}\chi_\mathrm{r}} \mathrm{.} \\
\end{equation}
\if\bibinc n
\bibliography{H2Orefs}
\fi

\if\bibinc y

\fi

\end{document}